\documentclass[aps,prd,twocolumn,preprintnumbers,superscriptaddress,amssymb,eqsecnum,showpacs,showkeyes,secnumarabic,graphics,floatfix,nofootinbib,tightenlines,longbibliography]{revtex4-1}
\usepackage{graphicx}
\usepackage{bm}
\usepackage{cancel}
\usepackage{epsfig}
\usepackage{dcolumn}
\usepackage{amsmath}
\usepackage{hhline}
\usepackage{enumerate}
\usepackage[utf8]{inputenc}
\usepackage{ulem}
\usepackage[dvipsnames]{xcolor}
\usepackage[breaklinks=true,colorlinks=true,
linkcolor=blue,urlcolor=Blue,citecolor=MidnightBlue,
bookmarks=true,bookmarksopenlevel=2]{hyperref}
\usepackage{bm}
\usepackage{amsmath}
\usepackage{amssymb}
\usepackage{longtable}

\usepackage{color, colortbl}
\definecolor{LightGreen}{rgb}{0.88,1,0.88}
\definecolor{LightCyan}{rgb}{0.88,1,1}
\definecolor{LightRed}{rgb}{1,0.85,0.85}

\begin{document}
\newcommand{\newc}{\newcommand}

\newcommand{\be}{\begin{equation}}
\newcommand{\ee}{\end{equation}}
\newcommand{\ba}{\begin{eqnarray}}
\newcommand{\ea}{\end{eqnarray}}
\newc{\D}{\partial}
\newc{\ie}{{\it i.e.} }
\newc{\eg}{{\it e.g.} }
\newc{\etc}{{\it etc.} }
\newc{\etal}{{\it et al.}}
\newc{\lcdm}{$\Lambda$CDM }
\newc{\lcdmnospace}{$\Lambda$CDM}
\newc{\plcdm}{Planck18/$\Lambda$CDM }
\newc{\plcdmnospace}{Planck18/$\Lambda$CDM}
\newcommand{\nn}{\nonumber}
\newc{\ra}{\Rightarrow}
\newc{\omm}{$\Omega_\mathrm{m,0}$ }
\newc{\ommnospace}{$\Omega_\mathrm{m,0}$}

\preprint{IFT-UAM/CSIC-21-98}

\title{Late-transition vs smooth \boldmath{$H(z)$} deformation models \\for the resolution of the Hubble crisis}

\author{George Alestas}\email{g.alestas@uoi.gr}
\affiliation{Department of Physics, University of Ioannina, GR-45110, Ioannina, Greece}
\author{David Camarena}\email{david.f.torres@aluno.ufes.br}
\affiliation{PPGCosmo, Universidade Federal do Espírito Santo, 29075-910, Vitória, ES, Brazil}
\author{Eleonora Di Valentino}\email{e.divalentino@sheffield.ac.uk}
\affiliation{School of Mathematics and Statistics, University of Sheffield, Hounsfield Road, Sheffield S3 7RH, United Kingdom}
\author{Lavrentios Kazantzidis}\email{l.kazantzidis@uoi.gr}
\affiliation{Department of Physics, University of Ioannina, GR-45110, Ioannina, Greece}
\author{Valerio Marra}\email{valerio.marra@me.com}
\affiliation{Núcleo de Astrofísica e Cosmologia \& Departamento de Física, Universidade Federal do Espírito Santo, 29075-910, Vitória, ES, Brazil}
\affiliation{INAF -- Osservatorio Astronomico di Trieste, via Tiepolo 11, 34131, Trieste, Italy}
\affiliation{IFPU -- Institute for Fundamental Physics of the Universe, via Beirut 2, 34151, Trieste, Italy}
\author{Savvas Nesseris}\email{savvas.nesseris@csic.es}
\affiliation{Instituto de F\'isica Te\'orica UAM-CSIC, Universidad Auton\'oma de Madrid, Cantoblanco, 28049 Madrid, Spain}
\author{Leandros Perivolaropoulos}\email{leandros@uoi.gr}
\affiliation{Department of Physics, University of Ioannina, GR-45110, Ioannina, Greece}

\date{\today}

\begin{abstract}
Gravitational transitions at low redshifts ($z_t<0.1$) have been recently proposed  as a solution to the Hubble and growth tensions. Such transitions would naturally lead to a transition in the absolute magnitude $M$ of type Ia supernovae (SnIa) at $z_t$ (Late $M$ Transitions -  $LMT$) and possibly in the dark energy equation of state parameter $w$ (Late $w-M$ Transitions -  $LwMT$). Here, we compare the quality of fit of this class of models to cosmological data, with the corresponding quality of fit of the cosmological constant model ($\Lambda$CDM) and some of the best smooth $H(z)$ deformation models ($w$CDM, CPL, PEDE). We also perform model selection via the Akaike Information Criterion and the Bayes factor. We use the full CMB temperature anisotropy spectrum data, the baryon acoustic oscillations (BAO) data, the Pantheon SnIa data, the SnIa absolute magnitude $M$ as determined by Cepheid calibrators and the value of the Hubble constant $H_0$ as determined by local SnIa calibrated using Cepheids.
We find that smooth $H(z)$ deformation models perform worse than transition models for the following reasons:
1) They have a worse fit to low-$z$ geometric probes (BAO and SnIa data);
2) They favor values of the SnIa absolute magnitude $M$ that are lower as compared to the value $M_c$ obtained with local Cepheid calibrators at $z<0.01$;
3) They tend to worsen the  $\Omega_\mathrm{m,0}-\sigma_\mathrm{8,0}$ growth tension.
We also find that the $w-M$ transition model ($LwMT$) does not provide a better quality of fit to cosmological data than a pure $M$ transition model ($LMT$) where $w$ is fixed to the $\Lambda$CDM value $w=-1$ at all redshifts. We conclude that the $LMT$ model has significant statistical advantages over smooth late-time $H(z)$ deformation models in addressing the Hubble crisis. 
\end{abstract}
\maketitle

\section{Introduction}
\label{sec:Introduction}

The scenario considered as the standard model in cosmology is the cosmological constant $\Lambda$ and cold dark matter (CDM) model, hereafter denoted as $\Lambda$CDM, as it is remarkably successful in fitting cosmological and astrophysical observations on a vast range of scales. However, this scenario is not a first principles theory, and it is based on unknown quantities (dark matter, dark energy and inflation), therefore can be considered as a low energy and large scales  approximation to a physical law, which has yet to be discovered. In this context, the observational problems in the estimates of the main cosmological parameters, see Refs.~\cite{DiValentino:2020vhf,DiValentino:2020zio,DiValentino:2020vvd,DiValentino:2020srs}, can hint towards the presence of deviations from the \lcdm scenario~\cite{Perivolaropoulos:2021jda,CANTATA:2021ktz}.

In particular, the most statistically significant  inconsistency is the well known Hubble constant {\it $H_0$ tension}, currently above the $4\sigma$ level (see \cite{Verde:2019ivm,Riess:2019qba,DiValentino:2020vnx,DiValentino:2021izs,Perivolaropoulos:2021jda,Shah:2021onj} and references therein). This tension refers to the disagreement between the value of $H_0$ estimated from the Planck satellite data~\cite{Planck:2018vyg}, assuming a \lcdm model, and the $H_0$ measured by the SH0ES collaboration~\cite{Riess:2020fzl}. However, there are many ways to obtain the Hubble constant value, and most of the early indirect estimates agree with Planck, as the Cosmic Microwave Background (CMB) ground telescopes~\cite{ACT:2020gnv,SPT-3G:2021eoc} or the Baryon Acoustic Oscillations (BAO) measurements~\cite{eBOSS:2020yzd}, while most of the late time measurements agree with SH0ES, even if obtained with different teams, methods or geometric calibrators~\cite{Soltis:2020gpl,Pesce:2020xfe,Kourkchi:2020iyz,Schombert:2020pxm,Blakeslee:2021rqi}. Finally, there are a few measurements that are in agreement with both sides, as the Tip of the Red Giant Branch~\cite{Freedman:2021ahq}, even if the re-analysis of~\cite{Anand:2021sum} shows a better consistency with the SH0ES value, or those based on the time delay~\cite{Birrer:2020tax}.

An additional challenge for the standard model is the {\it growth tension}. Dynamical cosmological probes favor weaker growth of perturbations than geometric probes in the context of general relativity and the \plcdm standard model at a level of about $3\sigma$ \cite{Hildebrandt:2016iqg,Nesseris:2017vor,Macaulay:2013swa,Kazantzidis:2018rnb,Skara:2019usd, Kazantzidis:2019dvk, Perivolaropoulos:2019vkb}. It would therefore be of particular interest to construct theoretical models that have the potential to simultaneously address both the $H_0$ and growth tensions. 

A wide range of theoretical models have been proposed as possible resolutions of the Hubble tension \citep{DiValentino:2021izs, Kazantzidis:2019dvk}. They can be divided in three broad classes:
\begin{itemize}
\item
``Early time" models  that recalibrate the scale of the sound horizon at recombination by modifying  physics during the pre-recombination epoch.  These models deform the Hubble expansion rate before recombination at $z>1100$ by introducing early dark energy \citep{Karwal:2016vyq,Poulin:2018cxd,Sakstein:2019fmf,Niedermann:2019olb,Hill:2020osr,Murgia:2020ryi,DAmico:2020ods,Gogoi:2020qif,Chudaykin:2020igl,Chudaykin:2020acu,Agrawal:2019lmo,Niedermann:2020dwg,Ye:2020btb,Lin:2019qug,Braglia:2020bym,Hill:2021yec,Chang:2021yog,Ye:2021iwa,Gomez-Valent:2021cbe,Jiang:2021bab,Karwal:2021vpk,Poulin:2021bjr}, extra neutrinos or relativistic species at recombination~\citep{Vagnozzi:2019ezj,Seto:2021xua,Carneiro:2018xwq,Gelmini:2019deq,Gelmini:2020ekg,DEramo:2018vss,Pandey:2019plg,Xiao:2019ccl,Nygaard:2020sow,Blinov:2020uvz,Binder:2017lkj,Choi:2019jck,DiValentino:2017oaw,Escudero:2019gvw,Arias-Aragon:2020qip,Blinov:2020hmc,Flambaum:2019cih,Anchordoqui:2021gji,Das:2021guu,Fernandez-Martinez:2021ypo,Feng:2021ipq,Ghosh:2021axu}, features in the primordial power spectrum~\cite{Hazra:2018opk,Keeley:2020rmo}, or evaporatig primordial black holes~\cite{Nesseris:2019fwr} etc. These models however, can alleviate but not fully solve the Hubble tension~\cite{Arendse:2019hev,Lin:2021sfs,Schoneberg:2021qvd}, and they tend to predict stronger growth of perturbations than implied by dynamical probes like redshift space distortion (RSD) and weak lensing (WL) data and thus they worsen the  growth tension \citep{Jedamzik:2020zmd}. This issue however is still under debate \citep{Smith:2020rxx}. 
\item
Late time deformations of the Hubble expansion rate $H(z)$ that assume a deformation of the best fit \plcdm $H(z)$ at late times. With the term ``deformation" we refer to a modification of the Planck/$\Lambda$CDM best fit form of $H(z)$ such that the new form of $H(z)$ not only tends to satisfy the local measurements of $H_0$ instead of the CMB best fit value, but also leads to an angular scale of the sound horizon that is consistent with the observed CMB peaks.
The analytical method for the construction of such ``deformed" $H(z)$ is described in Ref. \cite{Alestas:2020zol}. In this context, $H(z)$ retains its consistency with the observed CMB anisotropy spectrum while reaching the locally measured value of $H(z=0)$. In this class of models we can find both interacting dark matter~\citep{Hryczuk:2020jhi,Vattis:2019efj,Haridasu:2020xaa,Clark:2020miy,Mawas:2021tdx,Liu:2021mkv} or dark energy cosmologies~\citep{DiValentino:2019ffd,DiValentino:2019jae,Yang:2020uga,DiValentino:2020vnx,DiValentino:2020kpf,Yang:2021hxg,Anchordoqui:2021gji,Kumar:2017dnp,Kumar:2019wfs,Lucca:2020zjb,Yang:2019uog,Martinelli:2019dau,Gomez-Valent:2020mqn,DiValentino:2017iww,Kumar:2016zpg,Yang:2018euj,Pan:2020bur,Yao:2020hkw,Yao:2020pji,Pan:2019gop,Yang:2019uzo,Pan:2019jqh,Amirhashchi:2020qep,Gao:2021xnk,Lucca:2021eqy}, or extended and exotic dark energy models~\citep{Bertacca:2010mt,Vagnozzi:2019ezj,Haridasu:2020pms,Menci:2020ybl,Yang:2018qmz,DiValentino:2019dzu,DiValentino:2020naf,DiValentino:2020kha,Yang:2020ope,Benaoum:2020qsi,Yang:2021flj,DiValentino:2021zxy,Yang:2021eud,Li:2019yem,Pan:2019hac,Rezaei:2020mrj,Li:2020ybr,Hernandez-Almada:2020uyr,Banihashemi:2018oxo,Li:2019san,Yang:2020zuk,Banihashemi:2020wtb,Sola:2017znb,Keeley:2019esp,Dutta:2018vmq,daSilva:2020mvk,Guo:2018ans,daSilva:2020bdc,DiValentino:2019exe,Adler:2019fnp,Akarsu:2021fol,Benisty:2021gde,Mazumdar:2021yje,Shrivastava:2021hsu,Zhou:2021xov,Geng:2021jso}. While the interacting dark energy models need further investigations,\footnote{See, for example, Ref.~\citep{Nunes:2021zzi} for a study of the IDE models with SnIa data.} smooth $H(z)$ deformations due to the extended dark energy cosmologies have difficulty in fitting low $z$ cosmological distance measurements obtained by BAO and SnIa data \cite{Banerjee:2020xcn}. In addition, this class of models tends to imply a lower value of SnIa absolute magnitude $M$ than the value implied by Cepheid calibrators \cite{Alestas:2021xes}. Thus, this class of models cannot fully resolve the Hubble problem \citep{Benevento:2020fev,Alestas:2020mvb,Yang:2021flj,Theodoropoulos:2021hkk}, as demonstrated also in the present analysis for a few extended dark energy cosmologies.
\item
Late time transitions at a redshift $z_t\lesssim 0.01$ of the SnIa absolute magnitude $M$ have also been proposed as possible models that have the potential to resolve the Hubble tension \citep{Alestas:2020zol, Marra:2021fvf}. These models assume an abrupt transition of $M$ to a lower value (brighter SnIa at $z>z_t$) by $\Delta M\simeq -0.2$ mag. Such a reduction of $M$ could have been induced  by a fundamental physics transition of the effective gravitational constant $G_{\rm eff}$. This type of transition\footnote{A possible evolution of the absolute magnitude $M$ has also been recently investigated in Refs. \citep{Kazantzidis:2020tko, Sapone:2020wwz, Kazantzidis:2020xta, Dainotti:2021pqg}} 
could coexist with a transition of the dark energy equation of state $w$ from $w=-1$ at $z>z_t$ to a lower value at $z<z_t$ (phantom transition). This class of models could fully resolve the Hubble problem while at the same time address the growth tension by reducing the growth rate of cosmological perturbation due to the lower value of $G_{\rm eff}$ at $z>z_t$~\citep{Marra:2021fvf}.
Such models are highly predictive and have been challenged by existing, see Ref.~\cite{Sapone:2020wwz}, and upcoming (e.g.~Gravitational Waves Standard Sirens and Tully-Fisher data \cite{Alestas:2021nmi}) cosmological and astrophysical \cite{Alestas:2022xxm} data. Observationally, viable theoretical models that can support this transition include scalar-tensor theories with potentials where a first order late phase transition takes place \cite{EspositoFarese:2000ij,Ashoorioon:2014yua,Dainotti:2021pqg}.
\end{itemize}

Most previous studies usually marginalize over the SnIa absolute magnitude, treating it as a nuisance parameter~\cite{Verde:2009tu, SNLS:2011lii, SDSS:2014iwm, Pan-STARRS1:2017jku}. In particular, they consider the $M$-independent  $\bar{\chi}^2 \equiv -2\log \int \mathrm{d}M \exp(-\chi^2/2)$ function instead of the full $\chi^2$ function which explicitly  depends on $M$. In our analysis the parameter $M$ is not marginalized over and is included in the MCMC exploration along with the cosmological parameters. This allows us to compare our results with the corresponding inverse distance ladder constraints of \cite{Feeney:2017sgx,DES:2018rjw}, even though in the context of SnIa data its degeneracy with $H_0$ is acknowledged. However, in Refs.~\cite{Feeney:2017sgx,DES:2018rjw} the case of a transition in $M$ is not considered and this is one important difference from our approach along with the types of data considered in the fit. Thus, in view of the latter class of models, the designation $M$ tension/crisis might be more suitable to describe the problem \cite{Camarena:2021jlr,Efstathiou:2021ocp}. At its core, the issue is due to the fact that the supernova absolute magnitude $M$  used to derive the local $H_0$ constraint by the SH0ES collaboration is at a mismatch with the value of $M$ that is necessary to fit  SnIa,  CMB and  BAO data.

Note that the local distance ladder methodology of SH0ES considers SnIa data in the redshift range  of $0.023<z<0.15$. This makes the overall  method oblivious to any transitions in the value of $M$ at very low redshifts \cite{Alestas:2020zol}. In particular, the distance ladder methodology makes the crucial assumption that $M$ is the same at all redshifts. If this assumption is withdrawn and a transition is allowed at $z<0.01$ then the inferred value of $H_0$ may change significantly. For example if the transition occurs at $z=0.01$ then the calibrated value of $M$ correctly obtained at $z<0.01$ will not be the same as the value of $M$ at $z>0.01$ even though the value is assumed to be the same in the distance ladder methodology. Also, if the $M$ transition takes place at $z<0.01$ and the calibration analysis does not allow for such a transition then an incorrect value for $M$ will be obtained by the calibration analysis\cite{Perivolaropoulos:2021bds}. Since in the Hubble flow ($z>0.01$), the SnIa absolute magnitude $M$ is degenerate with $H_0$ through the observable ${\cal M} = M -5\,\log_{10}(h)+42.38$, where $h=H_0/100 \;\mathrm{km} \, \mathrm{s}^{-1}\,\mathrm{Mpc}^{-1}$, it becomes clear that if the true value of $M$ in the Hubble flow was lower than the value of $M$ for $z<0.01$ then correspondingly the true value of $H_0$ would also be lower and would become consistent with the CMB inferred value. If the transition occurs at $z < 0.01$, SnIa in the Hubble flow $(0.023 < z < 0.15)$ will naturally follow the calibration provided by CMB+BAO leading to a lower value on the Hubble constant.

Hints for such a late time transition may be seen in a recent re-analysis of the Cepheid SnIa calibration data where the Cepheid color-luminosity parameter is allowed to vary among galaxies \cite{Mortsell:2021nzg,Perivolaropoulos:2021bds}. More specifically, in Refs. \cite{Mortsell:2021nzg,Perivolaropoulos:2021bds} hints were found for a transition of this parameter or at least for it having a different value for the anchor galaxies compared to the SnIa host galaxies. Even though the errors of the individual $R_E$ parameters for each host are consistent with the corresponding anchor (low distance) values (see Figs. 4 and 5 of Ref. \cite{Mortsell:2021nzg}), when binning is implemented, the hints for a consistently different value (transition) becomes statistically more significant as shown in Ref. \cite{Perivolaropoulos:2021bds}. It is important to note here that the binning performed in Ref. \cite{Perivolaropoulos:2021bds} is not an anchor-calibrator binning, but one based on distance (low distance bin vs high distance bin separated by a critical distance $D_c$). The latter type of binning is justified and in this case led to a $2\sigma$ difference regarding the best fit values of $R_W$ and $M_W$ between the low distance bin and the high distance bin, at  $D_c\approx18 \mathrm{Mpc}$. This $2\sigma$ level of mismatch is clearly not statistically significant enough, but it is of interest to consider that if this degree of freedom is allowed (different values of $R_W$ and/or $M_W$ between high and low distance bins) the favored value of the best fit Hubble parameter becomes consistent with the CMB inferred value. This issue however is currently under debate and needs to be carefully interpreted.

\begin{figure*}[t!]
\centering
\includegraphics[width = 0.495 \textwidth]{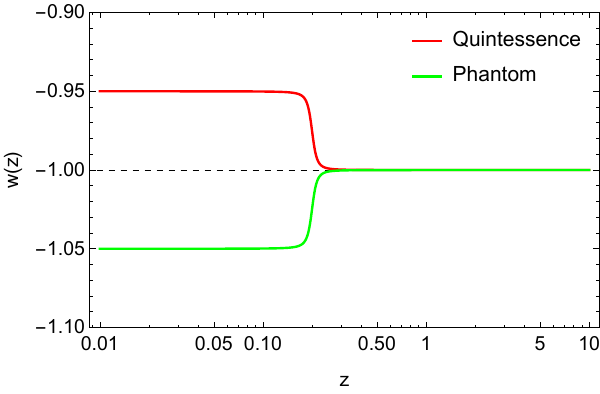}
\includegraphics[width = 0.495 \textwidth]{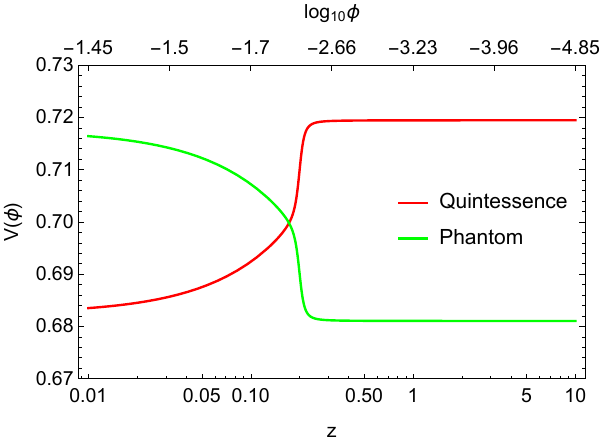}
\caption{An example of a transition in the dark energy equation of state $w$ (left panel) and how it can be caused by a sharp transition in a quintessence (red line) or phantom (green line) potential (right panel), with the scalar field running down/up the potential respectively. Here we assumed a smooth transition and reconstructed the potentials following Ref.~\cite{Sahni:2006pa}, assuming $\Omega_{\mathrm{m,0}}=0.3$, $z_t=0.2$, $\Delta w=\pm 0.05$ for quintessence/phantom fields.} 
\label{fig:transition}
\end{figure*}

In the present analysis we focus on late time $M$ transition models ($LMT$), possibly featuring also a transition in the dark energy equation of state parameter $w$ ($LwMT$), and compare their quality of fit to cosmological data with $H(z)$ deformation models. In particular, we address the following questions:
\begin{itemize}
\item 
How much does the quality of fit to low-$z$ cosmological data improve for $LMT$ models as compared to smooth $H(z)$ deformation models?
\item 
What is the level of $M$ transition favored by data? 
\item 
What is the value of $M$ favored by smooth $H(z)$ deformation models and how does it compare with the value of $M$ favored by Cepheid calibrators?
\item 
Does the addition of a $w$ transition on top of the $M$ transition significantly improve the quality of fit to the data?
\end{itemize}
Previous studies \cite{Marra:2021fvf,Alestas:2020zol,Alestas:2021xes} have indicated that $LMT$ models have improved quality of fit to cosmological data. However, those studies did not make use of the full CMB anisotropy spectrum but only effective parameters (shift parameter). The present analysis improves on those studies by implementing a more complete and accurate approach using the full Planck18 CMB anisotropy spectrum in the context of a Boltzmann code and a Monte Carlo Markov Chain (MCMC) analysis. 

The structure of our paper is the following: in the next Section~\ref{sec:lwmptdat} we focus on transition models ($LwMT$ and $LMT$) and present the constraints on their parameters using up-to-date cosmological data. In Section~\ref{sec:comparisonDE} we compare the quality of fit to cosmological data of transition models with the corresponding quality of fit of $H(z)$ deformation models; we also perform model selection. Finally, in Section~\ref{sec:concl} we summarize our results, discuss possible interpretations and present possible extensions of the present analysis.

\section{Transition models confronted by Observational Data \label{sec:lwmptdat}}

The $LMT$ model includes a sharp transition in the SnIa absolute magnitude $M$ of the form
\be 
M(z)=M_<+\Delta M \, \Theta(z-z_t), \label{eq:MLwMPT}
\ee
where $z_t$ is the transition redshift, $M_< \equiv M_c=-19.24$~mag is the local Cepheid-calibrated value from SH0ES as reconstructed in Refs.~\cite{Camarena:2019moy,Camarena:2021jlr} (in this Section we neglect uncertainties on $M_c$), $\Delta M$ is the parameter that quantifies the shift from the $M_c$ value, and $\Theta$ is the Heaviside step function. The $LwMT$ was first introduced in Ref.~\cite{Alestas:2020zol} and has, in addition to the $M$ transition, a dark energy equation of state $w$ transition of the form
\be 
w(z)=-1+\Delta w \, \Theta(z_t-z),
\ee
where $\Delta w$ describes the shift from the \lcdm value $(w=-1)$ for $z<z_t$.  Both $\Delta w$ and $\Delta M$ are parameters to be determined by the data.

\begin{figure*}[ht!]
\centering
\includegraphics[width = 0.98 \textwidth]{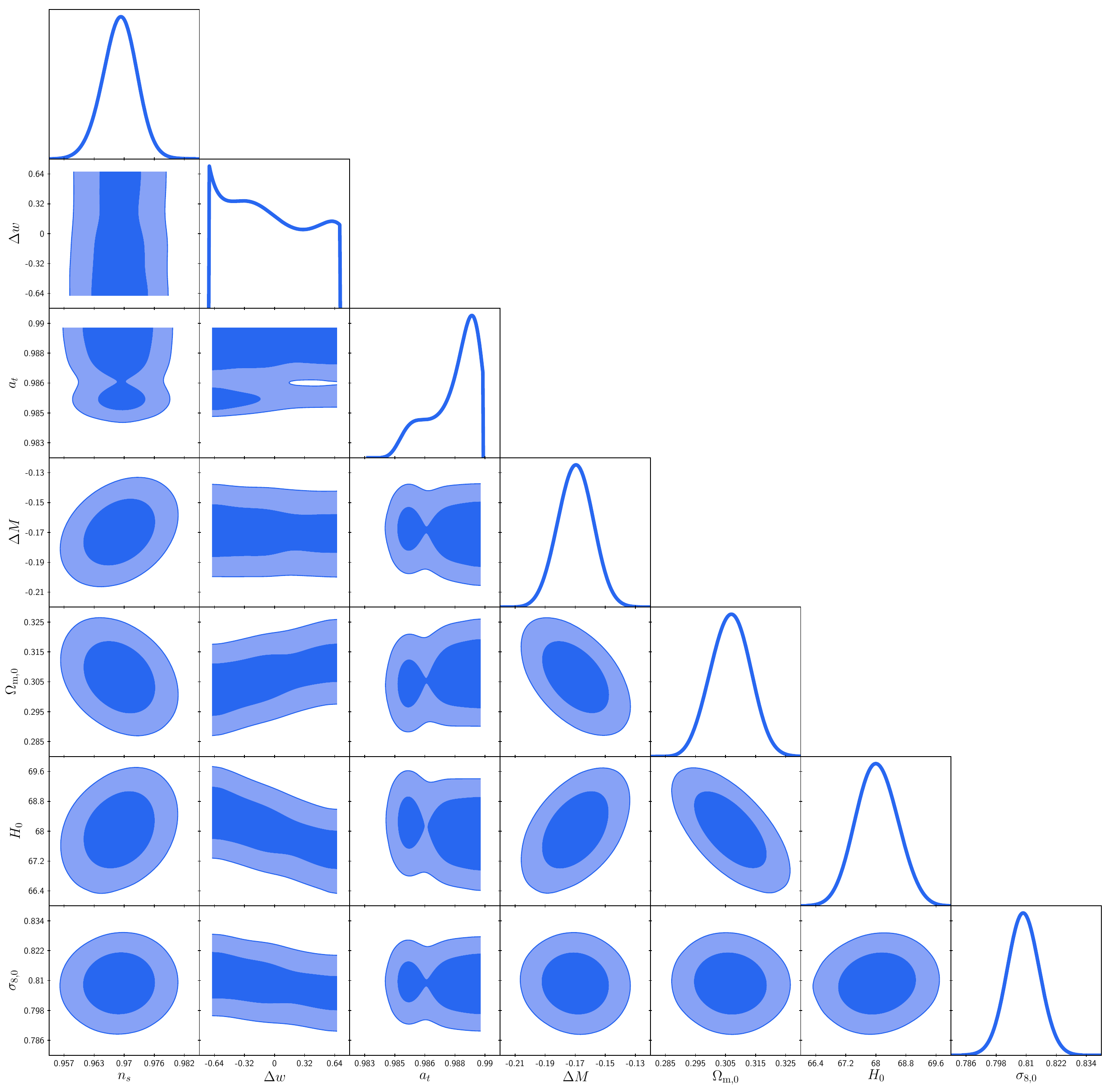}
\caption{The $68.3 \%$--$95.5 \%$ confidence contours for the parameters of the $LwMT$ model with $z_t \geq 0.01$, using the CMB+BAO+Pantheon+RSD likelihoods.} 
\label{fig:contours_LwMPT}
\end{figure*}

Such transitions in the dark energy equation of state $w(z)$ are in principle well-motivated and can easily happen within the context of a minimally coupled scalar field in general relativity (GR), either of the quintessence or phantom type. For example, in Fig.~\ref{fig:transition} we show a transition in the dark energy equation of state $w$ (left panel) and how it can be caused by a sharp transition in a quintessence (red line) or phantom (green line) potential (right panel), with the scalar field running down/up the potential respectively. For this plot we assumed, as an example, a smooth transition and reconstructed the potentials following the procedure of Ref.~\cite{Sahni:2006pa}, assuming $\Omega_{\mathrm{m,0}}=0.3$, $z_t=0.2$, $\Delta w=\pm 0.05$ for quintessence $(+)$ and phantom $(-)$ fields. By adjusting the aforementioned parameters, one may tune both the steepness and the redshift of the transition.

In order to constrain these transition models we use the following data combination:
\begin{itemize}
    \item The  Planck18 temperature CMB data, including the TTTEEE likelihoods for high-$l$ $(l>30)$, the temperature data TT and EE power spectra data for low-$l$ $(2<l<30)$~\citep{Aghanim:2018eyx,Planck:2019nip}, as well as the CMB lensing likelihood~\citep{Planck:2018lbu}.
    \item The BAO data presented in Refs.~\cite{Alam:2016hwk,Beutler:2011hx,Ross:2014qpa} as well as the Ly$\alpha$ BAO data of Refs.~\cite{Blomqvist:2019rah,Agathe:2019vsu}.
    \item The latest SnIa dataset (Pantheon) presented in Ref.~\cite{Scolnic:2017caz}.
    \item A robust redshift space distortion (RSD) compilation discussed in Ref.~\cite{Sagredo:2018ahx}, using the likelihood presented in Ref.~\cite{Arjona:2020yum}.
\end{itemize}
To analyze the data and obtain the best fit parameters we modify the publicly available \texttt{CLASS} code\footnote{For a step-by-step guide for the modifications implemented in \texttt{CLASS}, see \href{https://cosmology.physics.uoi.gr/wp-content/uploads/2021/07/Class_Implementation-1.pdf}{this file}.} and perform the Monte Carlo Markov Chain (MCMC) analysis using the publicly available \texttt{MontePython} code \cite{Brinckmann:2018cvx, Audren:2012wb, Blas_2011}. 

\begin{figure*}[ht!]
\centering
\includegraphics[width = 0.85 \textwidth]{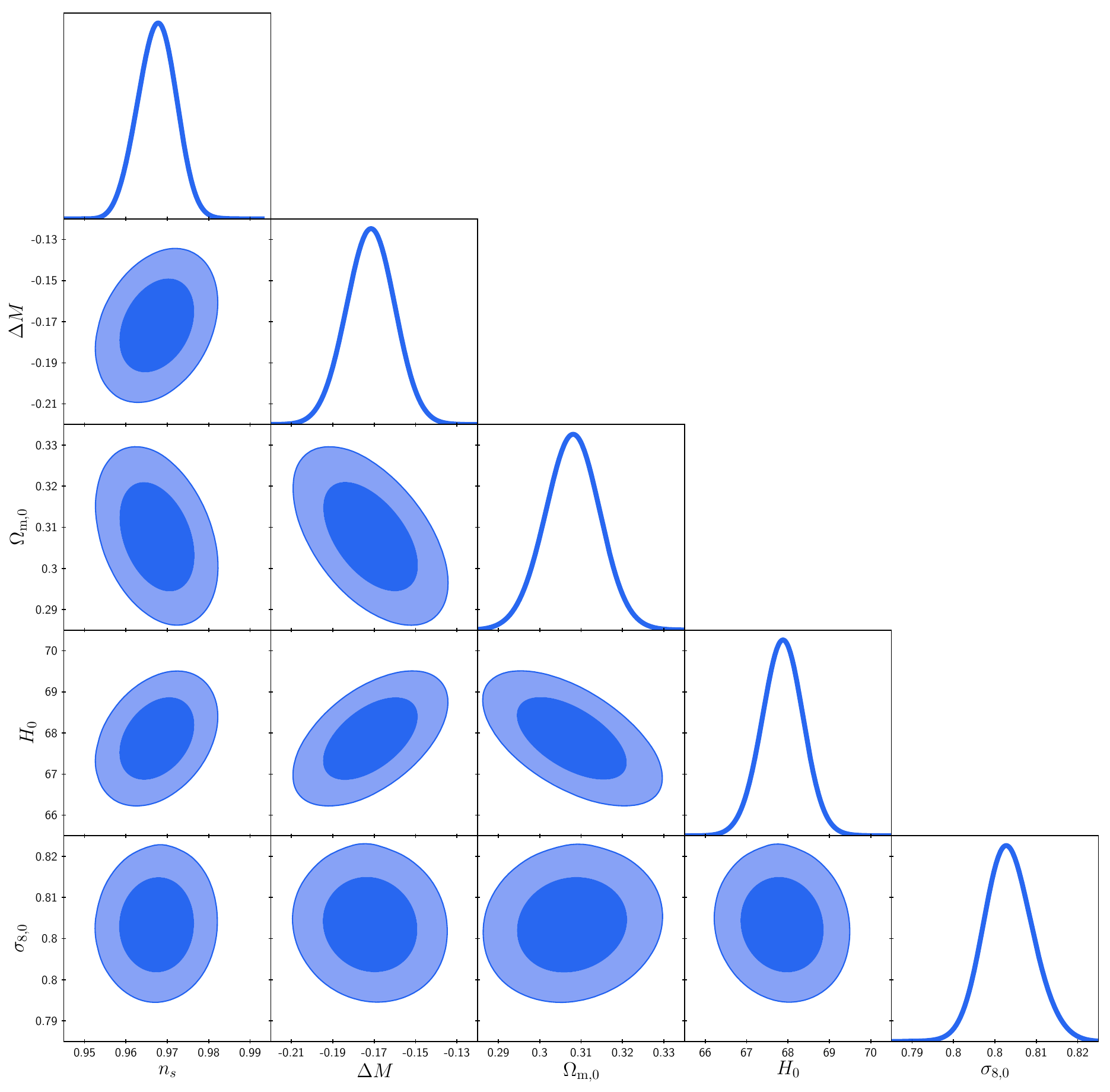}
\caption{The $68.3 \%$--$95.5 \%$ confidence contours for the $LMT$ model with $z_t=0.01$, using the CMB+BAO+Pantheon+RSD likelihoods.} 
\label{fig:contours_LMPT}
\end{figure*}   

These models by construction provide a great amount of flexibility in fitting the observational data since they can mimic \lcdm for $z>z_t$, while being fully consistent with local measurements of  $M$. In the case of the transition occurring at very low redshifts where there are almost no available data, \ie at $z_t<0.01$, we would normally anticipate a fit even better to that of \lcdm due to the extra parameter $\Delta w$ in the context of $LwMT$. However, then there would be no $H_0$ tension, since the local measurement of $H_0$ should coincide with the measurement of Planck if the $M$ transition is taken into account (a shift of $M$ implies a shift of $H_0$ since the two parameters are degenerate). 

Interestingly there are some works that use data with $z<0.01$, such as the extended Pantheon dataset of the latest SH0ES analysis (Panteon+) \cite{Riess:2021jrx} as well as the analyses of Refs. \cite{Dhawan:2020xmp,Perivolaropoulos:2021bds}, that can be used to search self-consistently for a transition in $M$ at $z<0.01$ using the combined Cepheid and SnIa data. Regarding the Pantheon+ dataset however, not only the data are not publicly available yet but also in our analysis we are simultaneously marginalizing over $M$ and $H_0$. We also stress that we are not including the full covariance between calibrators and supernovae as in the latest SH0ES analysis. Regarding Ref.~\cite{Dhawan:2020xmp}, the analysis makes no attempt to investigate an $M$ transition or to constrain variations of $H(z)$ for $z<0.01$ since this redshift region is not in the Hubble flow and thus it can not be reliably constrained. In contrast, it demonstrates that variations of the $H(z)$ parametrization in the Hubble flow (for $z>0.01$) do not affect the best fit value of $H_0$. This result could have been anticipated by the fact that (almost) all $H(z)$ parametrizations reduce to a cosmographic expansion in the range $0.01<z<0.1$ where the fit for $H_0$ is performed.

In addition, a separate analysis of Ref. \cite{Perivolaropoulos:2021bds}, focusing on the Cepheid+SnIa data for $z<0.01$, has found hints for a transition in the Cepheid absolute magnitude $M_W$ and in the color luminosity parameter $R_W$ which, if taken into account, make the value of absolute magnitude $M$ of SnIa consistent with its inverse distance ladder value, thus resolving the Hubble tension.

Hence, in what follows we impose a prior of $z_t \geq 0.01$ that corresponds to $a_t \leq 0.99$, since any lower value of $z_t$ cannot be probed via the considered Hubble flow data. Moreover, we use a prior of $\Delta w \in [-0.7,0.7]$. The best fit values of the $LwMT$ model with $z_t\geq 0.01$ are shown in Table \ref{tab:LwMPTbf}, while the $1\sigma-2 \sigma$ corresponding contours are shown in Fig.~\ref{fig:contours_LwMPT}. In Table \ref{tab:LwMPTbf} we also include the parameter $M_> \equiv M_c+\Delta M$ that arises for $z>z_t$ with $M_c$ corresponding to the Cepheid-calibrated value of the SnIa absolute magnitude.

\begin{table}[t!]
\caption{ The best-fit values and constraints at 68.3\% CL and 95.5\% CL of the parameters for the $LwMT$ model and $z_t \geq 0.01$ (or equivalently $a_t \leq 0.99$) using the CMB+BAO+Pantheon+RSD likelihoods described above. }
\scalebox{0.75}{                                              
\begin{tabular}{|c|c|c|c|c|} 
 \hline 
Parameter & best-fit & mean$\pm\sigma$ & 95.5\% lower & 95.5\% upper \\ 
\hline 
\rule{0pt}{3ex} 
$\Omega_\mathrm{m,0}$ &$0.3018$ & $0.3066_{-0.0065}^{+0.0064}$ & $0.2939$ & $0.3196$ \\
$n_s$ & $0.9708$ & $0.9685_{-0.0037}^{+0.0038}$ & $0.9608$ & $0.9759$ \\
$H_0$ &$68.56$ & $68.03_{-0.58}^{+0.55}$ & $66.94$ & $69.15$ \\ 
$\sigma_\mathrm{8,0}$ & $0.8141$ & $0.8089 \pm 0.0065$ & $0.7957$ & $0.8219$ \\
\rowcolor{LightGreen}
$\Delta M$ &$-0.1676$ & $-0.1698 \pm 0.012$ & $-0.1933$ & $-0.1467$ \\
$\Delta w$ & unconstrained & unconstrained & unconstrained & unconstrained \\ 
$a_t$ &$0.9856$ & $>0.985$ & $>0.984$ & $>0.984$ \\ 
\rowcolor{LightGreen}
$M_> \equiv M_c+\Delta M$ & $-19.408$ & $-19.410 \pm 0.012$ & $-19.433$ & $-19.387$\\
\hline \hline
$-\ln{\cal L}_\mathrm{min}$ & \multicolumn{4}{|c|}{$1917.02$}  \\
$\chi^2_\mathrm{min}$ & \multicolumn{4}{|c|}{$3834$} \\
\hline
 \end{tabular}
}                                                                                   
\label{tab:LwMPTbf}                                                                 \end{table}

\begin{table}[t!]
\caption{The best-fit values and constraints at 68.3\% CL and 95.5\% CL of the parameters for the $LMT$ model with $z_t=0.01$ (or equivalently $a_t=0.99$) using the CMB+BAO+Pantheon+RSD likelihoods.}
\scalebox{0.85}{  
\begin{tabular}{|c|c|c|c|c|} 
 \hline 
Parameter & best-fit & mean$\pm\sigma$ & 95.5\% lower & 95.5\% upper \\ 
\hline 
\rule{0pt}{3ex} 
$\Omega_\mathrm{m,0}$ & $0.3088$ & $0.3082_{-0.0058}^{+0.0052}$ & $0.2976$ & $0.3193$ \\
$n_s$ & $0.9697$ & $0.968_{-0.0037}^{+0.0038}$ & $0.9606$ & $0.9754$ \\
$H_0$ & $67.88$ & $67.89_{-0.40}^{+0.42}$ & $67.06$ & $68.71$ \\ 
$\sigma_\mathrm{8,0}$ & $0.8085$ & $0.8084_{-0.0061}^{+0.0058}$ & $0.7963$ & $0.8205$ \\
\rowcolor{LightGreen}
$\Delta M$ & $-0.170$ & $-0.172 \pm 0.012$ & $-0.195$ & $-0.149$ \\ 
\rowcolor{LightGreen}
$M_> \equiv M_c+\Delta M$ & $-19.410$ & $-19.412 \pm 0.012$ & $-19.435$ & $-19.389$\\
\hline \hline
$-\ln{\cal L}_\mathrm{min}$ & \multicolumn{4}{|c|}{$1917.52$}  \\
$\chi^2_\mathrm{min}$ & \multicolumn{4}{|c|}{$3835$} \\
\hline
 \end{tabular}
}                                                                                   
\label{tab:LMPTbf}                                                                 \end{table}   

From Table \ref{tab:LwMPTbf}, we see that the parameter $a_t$ (or equivalently $z_t$) approaches the highest (lowest) value imposed by the data in order to achieve the best possible quality of fit favoring a transition at very low redshifts.  Moreover, the posterior probability of $a_t$ appears to be bimodal. The reason for this behavior may be seen \eg in Fig. 9 of Ref. \cite{Alestas:2020zol} or in Fig. 3 of Ref. \cite{Camarena:2021jlr} where the lowest $z$ bin for the SnIa absolute magnitude $M$ shows a rise that may be interpreted as a hint for a transition at $z  \simeq 0.015$ (this is expressed by the first plato-peak for the likelihood of $a_t$ at $a \simeq 0.0986$). The higher peak however occurs at $a_t=0.99$ indicating that the minimum $\chi^2$ for the transition redshift is at or below $z=0.01$ (no clear hint for a transition in the Hubble flow). However, given that the data do not extend to more recent times than $a=0.99$, the best fit is at $a_t=0.9856$ as it is explicitly written in Table \ref{tab:LwMPTbf} and the higher peak at $a_t =0.99$ can only be interpreted as a lower bound for the value $a_t$ if the transition occurs at more recent times. We thus chose to neglect this second peak.

The timing of the transition is not particularly fine-tuned due to the fact that at very low redshifts dark energy has started to dominate in the Universe. Since at that time $\Omega_{\Lambda}>0.5$, new physics could possibly emerge. Furthermore, in previous analyses by some of the authors of the current work \cite{Kazantzidis:2020tko, Kazantzidis:2020xta} a tomographic analysis of the Pantheon dataset has been performed. In both of these references, it has been shown that for $z>0.01$ the redshift binned best fit \lcdm parameter values for the parameter $M$ (as well as $\Omega_{m,0}$) vary around the full dataset fit value (assumed constant) by up to $\Delta M=0.08\pm 0.06$. This variation is significantly smaller than the variation $\Delta M\simeq 0.2$ required for the resolution of the Hubble tension (see \eg the left panel of Fig. 1 of \cite{Kazantzidis:2020xta}). Most importantly, however, we observe that despite allowing for an extra degree of freedom induced by having $\Delta w\neq 0$, this parameter seems to be ignored by the data. 

Since $z_t \approx 0.01$ is favored by the data, the parameter $\Delta w$ becomes irrelevant due to the fact that for $z_t=0.01$, $\Delta w$ would modify the expansion rate $H(z)$ only in a region where there are no data available ($z<0.01$). This carries the implication that a $w$ transition is perhaps not needed in order to obtain the best quality of fit to the data\footnote{The $w$ transition however may be required for theoretical reasons. In scalar tensor theories a gravitational transition to weaker gravity at early times may require a simultaneous transition to $w<-1$ at late times.}. We thus repeat the analysis considering only an $M$ transition (``Late $M$ Transition" - $LMT$), setting $\Delta w=0$ and $a_t=0.99$ (or equivalently $z_t=0.01$), which is basically the maximum of the posterior of $a_t$ for the $LwMT$ model.
We obtain the best fit and mean values as indicated in Table \ref{tab:LMPTbf}; the  contours are shown in Fig.~\ref{fig:contours_LMPT}.

As we can see comparing Tables \ref{tab:LwMPTbf}, \ref{tab:LMPTbf} and Figs.~\ref{fig:contours_LwMPT}, \ref{fig:contours_LMPT} the introduction of $\Delta w$ has practically no effect on the quality of fit, \ie on the $\chi^2$ value.
Moreover, the mismatch between the local calibration of the SnIa absolute magnitude and the value inferred from the other probes is very significant, suggesting that the designation $M$ tension/crisis is  suitable to describe the $H_0$ crisis \cite{Camarena:2021jlr,Efstathiou:2021ocp}.
Finally, it is interesting to note that the inferred value of $M_>=-19.41$ mag agrees well with the constraint $M=-19.40$ mag that was obtained using the parametric-free inverse distance ladder of Ref.~\cite{Camarena:2019rmj}.

So, some natural questions that arise are the following: ``How do the transition models $LwMT$ and $LMT$ compare to some other popular dark energy models in the literature that also try to address the Hubble tension?" and ``Can these models provide an $M$ value that is consistent with the Cepheid measurement $M_c$ as the transition models that we discussed?" These questions will be addressed in the following section, where we perform a comparison between some popular dark energy parametrizations (smooth deformation dark energy models) with the transition models $LwMT$ and $LMT$.

\begin{centering}
\begin{table*}[t]
\caption{Constraints at 68.3\% CL of the cosmological parameters for all the dark energy models explored in this work when a narrow flat prior $M \in [-19.28,-19.2]$ mag is assumed, forcing the agreement with Cepheid calibration \cite{Camarena:2019moy,Camarena:2021jlr} at the $1\sigma$ level.
Note that this prior is artificial and the correct prior is the Gaussian one of Eq.~\eqref{Mprior} which is adopted in Table~\ref{tab:bfall_gaussian}.
For the transition models, $M$ is fixed to $-19.24 \text{ mag } (M_<)$. $\Delta \chi^2$ corresponds to the $\chi^2_\mathrm{min}$ difference of each model with the \lcdm case. All models provide a much better overall fit as compared to $\Lambda$CDM, and the $LwMT$ and $LMT$ models fair considerably better than the rest.}
\label{tab:bfall}
\resizebox{\textwidth}{!}{
\begin{tabular}{|c|c|c|c|c|c|c|}
 \hline 
 \rule{0pt}{3ex}  
  Parameters & \lcdm & $w$CDM & CPL & $LwMT$ & PEDE & $LMT$\\
  & & & & $(z_t \geq 0.01)$ & & $(z_t=0.01)$\\
    \hline
    \rule{0pt}{3ex}  
$\Omega_\mathrm{m,0}$ & $0.2564^{+0.0018}_{-0.0019}$ & $0.2571^{+0.0019}_{-0.0020}$ & $0.2719^{+0.0041}_{-0.0044}$ & $0.3066 \pm {0.0063}$ & $0.2582 \pm {0.0020}$ & $0.3082 \pm {0.0053}$ \\ 
$n_s$ & $0.992 \pm {0.003}$ & $0.972 \pm {0.004}$ & $0.967 \pm {0.004}$ & $0.968 \pm {0.004}$ & $0.971 \pm {0.003}$ & $0.968 \pm {0.004}$ \\ 
$H_0$ & $72.40 \pm {0.16}$ & $73.99^{+0.26}_{-0.27}$ & $72.38 \pm {0.48}$ & $68.03 \pm {0.55}$ & $73.90^{+0.17}_{-0.19}$ & $67.89 \pm {0.40}$ \\ 
$\sigma_\mathrm{8,0}$ & $0.8045^{+0.0072}_{-0.0081}$ & $0.8507^{+0.0084}_{-0.0083}$ & $0.8511^{+0.0084}_{-0.0081}$ & $0.8088 \pm {0.0063}$ & $0.8517 \pm {0.0059}$ & $0.8084 \pm {0.0059}$ \\ 
$S_\mathrm{8}$ & $0.7437 \pm {0.0077}$ & $0.7876 \pm {0.0084}$ & $0.8103 \pm {0.0100}$ & $0.8177^{+0.0101}_{-0.0103}$ & $0.7901 \pm {0.0065}$ & $0.8194 \pm {0.0100}$ \\ 
\rowcolor{LightGreen}
$M$ & $\sim -19.28$ & $\sim -19.28$& $\sim -19.28$ & $-19.24\, (M_<)$ & $\sim -19.28$ & $-19.24 \, (M_<)$ \\ \hline 
\cellcolor{LightGreen}$\Delta M$ & - & - & - & \cellcolor{LightGreen}$-0.170 \pm {0.011}$ & - & \cellcolor{LightGreen}$-0.172 \pm {0.011}$ \\ 
\cellcolor{LightGreen}$M_> \equiv M_c+\Delta M$ & - & - & - & \cellcolor{LightGreen}$-19.410 \pm {0.011}$ & - & \cellcolor{LightGreen}$-19.412 \pm {0.011}$ \\ 
$\Delta w$ & - & - & - & unconstrained & - & - \\ 
$a_t$ & - & - & - & $>0.987$ & - & - \\ 
$w_0$ & - & $-1.162^{+0.021}_{-0.019}$ & $-0.844^{+0.077}_{-0.089}$ & - & - & - \\ 
$w_a$ & - & - & $-1.27^{+0.38}_{-0.31}$ & - & - & - \\ 
\hline 
 \rule{0pt}{3ex}
$\chi^2_{\rm{min}}$ & $3964$ & $3889$ & $3875$ & $3834$ & $3886$ & $3835$ \\
\rowcolor{LightCyan}
$\Delta \chi^2_M$  & - & $-75$ & $-89$ & $-130$ & $-78$ & $-129$\\ \hline

\end{tabular}}
\end{table*}
\end{centering}

\section{Comparison of Dark Energy Models}
\label{sec:comparisonDE}
In order to truly resolve the $H_0$ tension, a dark energy model should not only provide a consistent measurement for $M$, but also maintain a quality of fit comparable (or even better) to \lcdm with low-$z$ data (BAO and SnIa), as discussed earlier. In this section, we consider some popular dark energy models, that have been suggested as being capable of addressing the $H_0$ tension, following three different methods:
\begin{enumerate}
    \item Force all the models to be consistent with the Cepheid absolute magnitude measurement \cite{Camarena:2019moy, Camarena:2021jlr} at the $1\sigma$ level by imposing a flat prior $M~\in~[-19.28,-19.2]$ mag.
    \item Analyze all models including the  local Cepheid-calibrated prior by SH0ES \citep{Camarena:2021jlr}:
    \begin{align} \label{Mprior}
        M_c=-19.24 \pm 0.04 \text{ mag.}
    \end{align}
    \item Include the SH0ES determination of $H_0$ \cite{Riess:2020fzl}, allowing at the same time the absolute magnitude $M$ to vary freely (flat prior). This is illustrated in Appendix~\ref{appA}, as a complementary analysis.
\end{enumerate}

\begin{figure*}[ht!]
\centering
\includegraphics[width =  \textwidth]{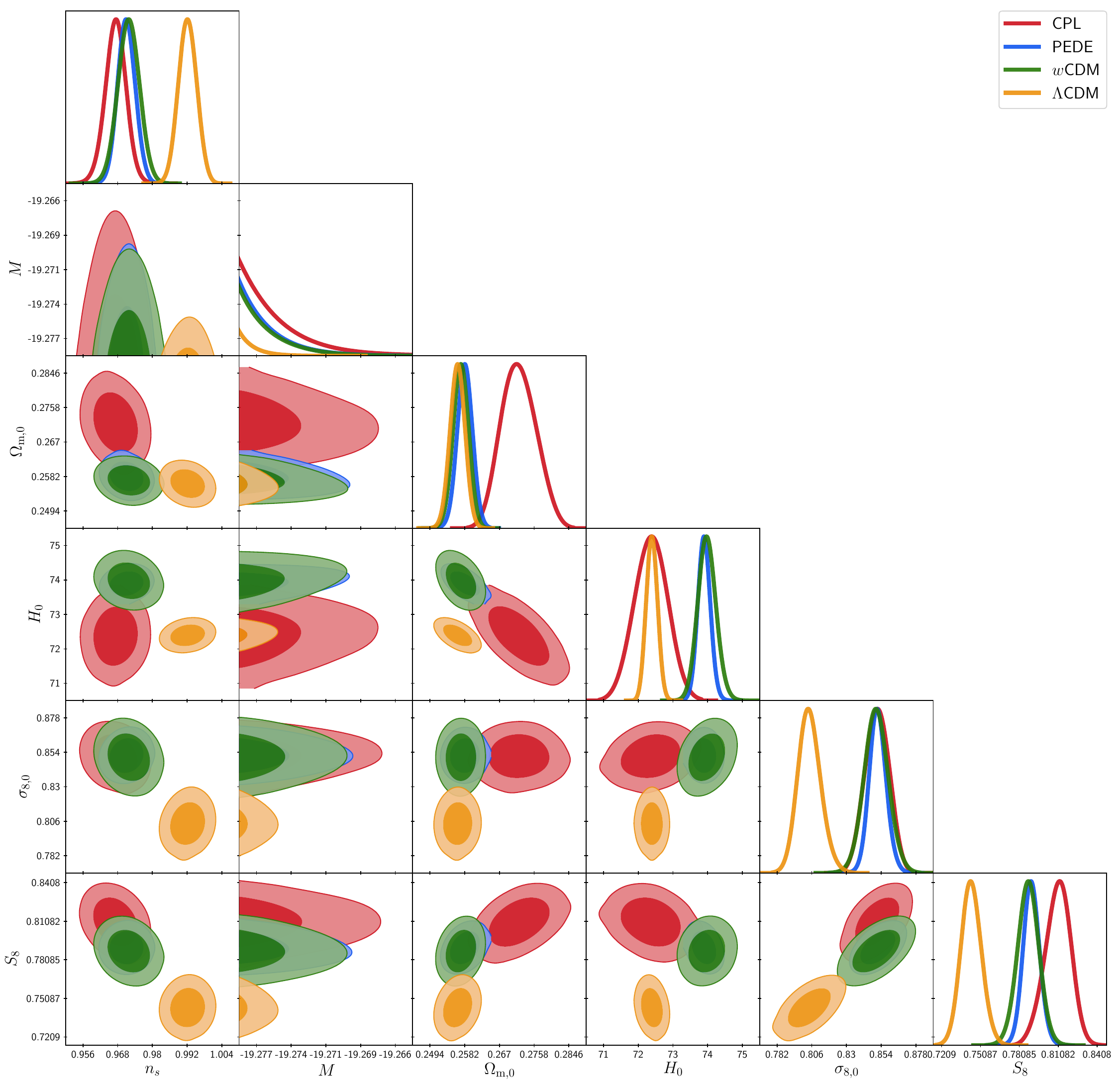}
\caption{The $68.3 \%$--$95.5 \%$ confidence contours for the common parameters of the $\Lambda$CDM, CPL, $w$CDM and PEDE dark energy models corresponding to the bounds illustrated in Table \ref{tab:bfall}. We used the CMB+BAO+Pantheon+RSD likelihoods, imposing the narrow flat prior $M \in [-19.28,-19.2]$ mag. The $M$ prior severely constrains the best fit of $M$ to the lowest possible value, displaying their tendency to provide a significantly lower value for $M$.} 
\label{fig:contours_Mpr1}
\end{figure*} 

The $H(z)$ deformation dark energy models that we consider in this work include the $w$CDM model, \ie a model with a constant equation of state $w$, assuming a flat Universe and cold dark matter, that is described by a Hubble parameter of the form (neglecting radiation and neutrinos at late times)
\be 
H(z)\!=\!H_0  \sqrt{\Omega_\mathrm{m,0}(1\!+\!z)^3\!+\!(1-\Omega_\mathrm{m,0})(1\!+\!z)^{3(1\!+\!w)}},
\ee
which for $w=-1$ reduces to the usual Hubble parameter for the \lcdm model. Moreover, we consider the Chevallier-Polarski-Linder (CPL) parametrization, with a dark energy equation of state \cite{Chevallier:2000qy,Linder:2002et}\textbf{}
\be 
w(z) = w_0 + w_{a} \left(\frac{z}{1+z} \right), \label{eq:wcpl}
\ee 
where $w_0$ and $w_a$ are free parameters. The corresponding Hubble parameter for the CPL model is the following

\be
H(z)=H_0 \, \sqrt{\begin{aligned}\Omega_\mathrm{m,0} &(1+z)^3 +\left(1-\Omega_\mathrm{m,0}\right) \times \\ & 
(1+z)^{3(1+w_0+w_a)}  e^{-3 \frac{w_a z}{1+z}}\end{aligned}}\,. \label{hzcpl}
 \ee

Furthermore, we consider the phenomenologically emergent dark energy (PEDE) model which shows significant promise in resolving the $H_0$ problem. This model was introduced in Ref. \cite{Li:2019yem} and has an equation of state of the form
\be 
w(z) = -\frac{1}{3\ln10}(1+\tanh[\log_{10}(1+z)]) - 1,  \label{eq:wpede}
\ee
with a corresponding Hubble parameter of the form

\be 
H(z)=H_0 \,\sqrt{\begin{aligned}\left(1-\Omega_\mathrm{m,0}\right) & \times \left[ 1 - {\tanh}\left(\log_{10}(1+z) \right) \right] \\
& + \Omega_\mathrm{m,0} (1+z)^3. \end{aligned}}
\ee

The main advantage of the aforementioned parametrization is that it has the same number of degrees of freedom as \lcdmnospace. Finally, we consider the transition models $LwMT$ with $z_t>0.01$ and $LMT$ with $z_t=0.01$ described in Sec.~\ref{sec:lwmptdat}, as well as the \lcdm model itself, thus having a total of six different models.

\begin{centering}
\begin{table*}[t]
\caption{Constraints at 68.3\% CL of the cosmological parameters for the dark energy models explored in this work when the prior $ M = -19.24 \pm 0.04$ mag of Eq.~\eqref{Mprior} from SH0ES is adopted. $\Delta \chi^2$ corresponds to the $\chi^2_\mathrm{min}$ difference of each model with the \lcdm case. Only transitions models provide a competitive fit to data as compared to $\Lambda$CDM.}
\label{tab:bfall_gaussian}
\resizebox{\textwidth}{!}{
\begin{tabular}{|c|c|c|c|c|c|c|}
\hline 
\rule{0pt}{3ex}  
Parameters & $\Lambda$CDM & $w$CDM & CPL & $LwMT$ & PEDE & $LMT$\\
& & & & $(z_t \geq 0.01)$ & & $(z_t=0.01)$\\
\hline
\rule{0pt}{3ex}  
$\Omega_\mathrm{m,0}$ & $0.3022_{-0.0052}^{+0.0051}$ & $0.2943 \pm {0.0065}$ & $0.2974^{+0.0067}_{-0.0068}$ & $0.3073^{+0.0063}_{-0.0062}$ & $0.2789 \pm {0.0049}$ & $0.3082 \pm {0.0053}$ \\ 
$n_s$ & $0.9704 \pm 0.004$ & $0.968 \pm {0.004}$ & $0.967 \pm {0.004}$ & $0.968 \pm {0.004}$ & $0.963 \pm {0.003}$ & $0.968 \pm {0.004}$ \\ 
$H_0$ & $68.36 \pm 0.4$ & $69.47 \pm {0.72}$ & $69.25 \pm {0.73}$ & $67.96 \pm {0.55}$ & $71.85 \pm {0.45}$ & $67.89 \pm {0.40}$ \\ 
$\sigma_\mathrm{8,0}$ & $0.8076_{-0.0062}^{+0.0058}$ & $0.8215^{+0.0095}_{-0.0097}$ & $0.8248^{+0.0096}_{-0.0097}$ & $0.8084^{+0.0064}_{-0.0065}$ & $0.8531 \pm {0.0059}$ & $0.8085 \pm {0.0057}$ \\ 
$S_\mathrm{8}$ & $0.8105_{-0.01}^{+0.0097}$ & $0.8135 \pm {0.0098}$ & $0.8210^{+0.0107}_{-0.0106}$ & $0.8181 \pm {0.0100}$ & $0.8226 \pm {0.0095}$ & $0.8194 \pm {0.0099}$ \\ 
\rowcolor{LightGreen}$M$ & $-19.40 \pm {0.01}$ & $-19.38 \pm {0.02}$ & $-19.37 \pm {0.02}$ & $-19.26 \pm {0.04}$ & $-19.33 \pm {0.01}$ & $-19.24 \pm {0.04}$ \\ \hline
\cellcolor{LightGreen}$\Delta M$ & - & - & - & \cellcolor{LightGreen}$-0.145^{+0.038}_{-0.035}$ & - & \cellcolor{LightGreen}$-0.168 \pm {0.039}$ \\ 
$\cellcolor{LightGreen}M_>$ & - & - & - & \cellcolor{LightGreen}$-19.410 \pm {0.011}$ & - & \cellcolor{LightGreen}$-19.411 \pm {0.011}$ \\ 
$\Delta w$ & - & - & - & unconstrained & - & - \\ 
$a_t$ & - & - & - & $> 0.986$ & - & - \\ 
$w_0$ & - & $-1.050 \pm {0.027}$ & $-0.917 \pm {0.078}$ & - & - & - \\ 
$w_a$ & - & - & $-0.53^{+0.33}_{-0.28}$ & - & - & - \\ 
\hline
\rule{0pt}{3ex}  
$\chi^2_{\rm{min}}$ & $3854$ & $3851$ & $3848$ & $3833$ & $3867$ & $3835$ \\
\rowcolor{LightCyan}
$\Delta \chi^2$ & - & $-3$ & $-6$ & $-21$ & $+13$ & $-19$ \\
\hline
\end{tabular}}
\end{table*}
\end{centering}

\begin{figure*}[t!]
\centering
\includegraphics[width = \textwidth]{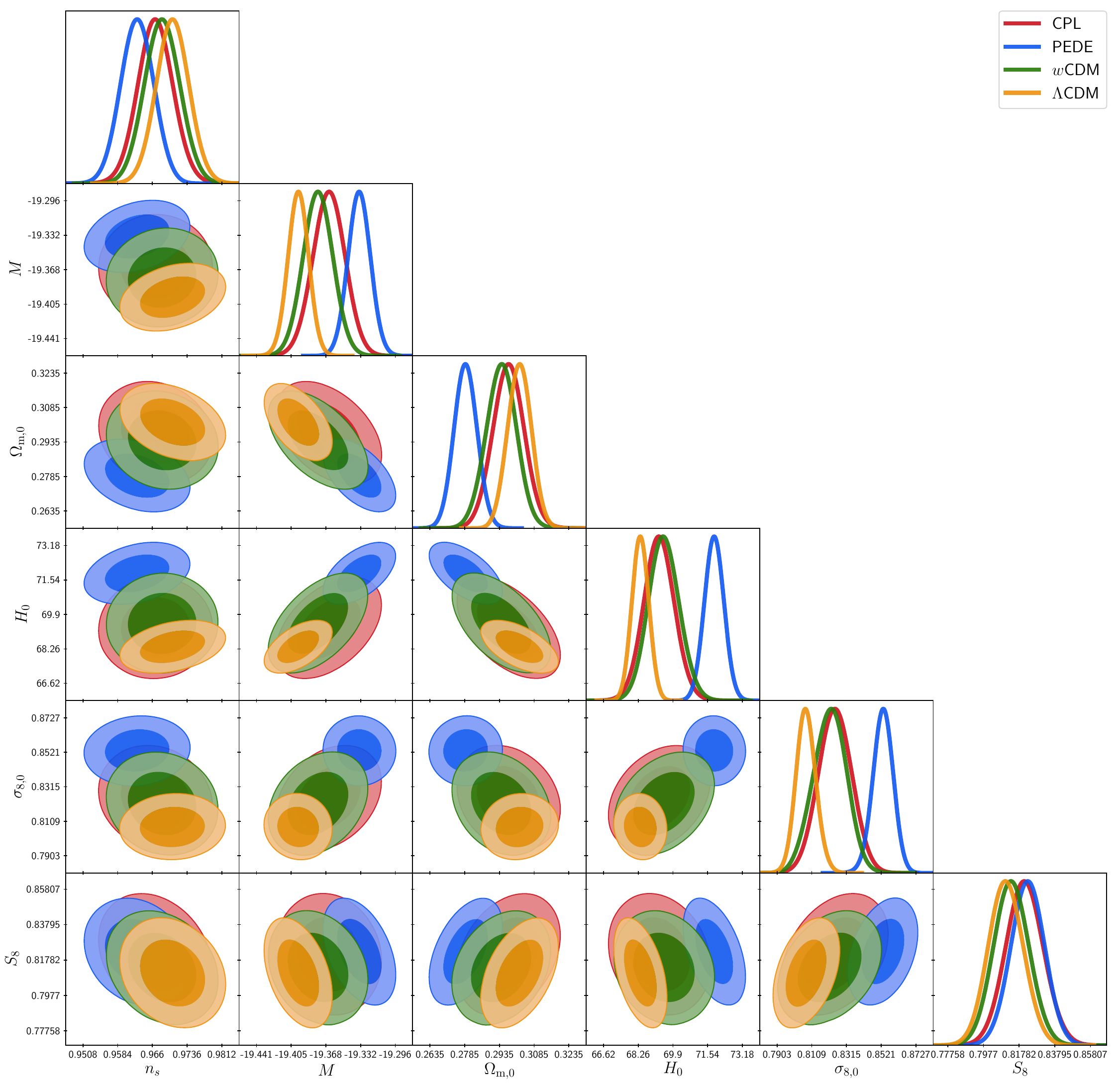}
\caption{The $68.3 \%$--$95.5 \%$ confidence contours for the common parameters of the CPL, $w$CDM  and PEDE dark energy models with the  prior $ M = -19.24 \pm 0.04$ mag of Eq.~\eqref{Mprior} from  SH0ES (corresponding bounds in Table \ref{tab:bfall_gaussian}).} 
\label{fig:contours_Mgaussian}
\end{figure*} 

\begin{centering}
\begin{table}[t]
\renewcommand{\arraystretch}{1.3}
\setlength{\tabcolsep}{9pt}
\caption{$\Delta \chi^2$ and  corresponding $\Delta AIC$ and $ \ln B$ values for all models of Table \ref{tab:bfall_gaussian} with respect to \lcdmnospace. Negative values of $\Delta \chi^2$ and $\Delta AIC$ and positive values of $ \ln B$ signal that a model is favored with respect to  \lcdmnospace.}
\label{tab:statcritgau}
\begin{tabular}{|c|c|c|c|}
 \hline 
 \rule{0pt}{3ex}  
  Gaussian $M$ Prior Case & $\Delta \chi^2$ & $\Delta AIC$ & $ \ln B$\\
    \hline
    \rule{0pt}{3ex}    
 \lcdm & $-$ & $-$ & $-$\\
  $LMT (z_t=0.01)$ & $-19$ & $-17$ & $+9.1$ \\
  $LwMT (z_t \geq 0.01)$ & $-21$ & $-15$ & $+6.2$ \\
  $w$CDM & $-3$ & $-1$ & $+2.2$ \\
  CPL & $-6$ & $-2$ & $-2.4$ \\
  PEDE & $+13$ & $+13$ & $-6.5$ \\
\hline
\end{tabular}
\end{table}
\end{centering}

\subsection{Dark energy models comparison using a narrow flat prior on \boldmath{$M~\in~[-19.28,-19.2]$} mag}

We perform the MCMC analysis using the likelihoods described in Section~\ref{sec:lwmptdat} and imposing a narrow flat prior on the SnIa absolute magnitude  $M \in [-19.28,-19.2]$~mag, that is, forcing  all models to be consistent with the Cepheid measurement $M_c$.
Rigorously, this prior is artificial as the correct prior is the Gaussian one of Eq.~\eqref{Mprior}.
However, the use of this narrow prior will be useful to understand the impact of the local calibration on the quality of fit of the various models.

For the transition models $LwMT$ and $LMT$, we use Eq.~\eqref{eq:MLwMPT} for the SnIa absolute magnitude and leave $\Delta M$ as a free variable. Thus, these are the only models that can, by construction, escape from the imposed $M$ prior. The constraints on the cosmological parameters as well as the $68.3 \%$--$95.5 \%$ confidence contours of the corresponding parameters of the models are shown in Table \ref{tab:bfall} and Fig. \ref{fig:contours_Mpr1} respectively. For the sake of clarity, contour plots for the $LMT$ and $LwMT$ models are displayed separately. In this case, the analysis with a narrow flat prior on $M$ produces same constraints as those showed as Fig.~\ref{fig:contours_LMPT} and Fig.~\ref{fig:contours_LwMPT} for the $LMT$ and $LwMT$ models, respectively. A wider flat prior would lead (as in the case of a Gaussian prior of Table \ref{tab:bfall_gaussian} that follows) the best fit value of $M$ for most models to be very close to the CMB inferred value of $M=-19.4$ with an error bar which makes it inconsistent with the Cepheid inferred vale $M=-19.24$. In view of the degeneracy of $M$ with $H_0$, this $M$ tension is closely related with the $H_0$ tension. A similar (but milder) tension occurs also in the case of a Gaussian prior on $M$ as indicated in Table \ref{tab:bfall_gaussian}.

All models, except the $LwMT$ with $z_t \geq 0.01$ and $LMT$ with $z_t=0.01$, give an $H_0$ value that is consistent with the SH0ES determination of $H_0$ \cite{Riess:2020fzl} and $M \sim -19.28\, \mathrm{mag}$, \ie the lowest eligible value of the prior that we imposed, displaying their tendency to provide a significantly lower value for $M$. On the other hand, the transition models provide a $H_0$ value close (within the $1\sigma$ level) to the typical \plcdm value, providing at the same time $M \approx -19.4 \, \mathrm{mag}$ as expected.

Note that the $\Lambda$CDM model has a very bad fit to the data as compared to $w$CDM, CPL and PEDE. This is due to the fact that, having fixed $M$ to the local $M_c$ value, supernova data constrain the $\Lambda$CDM model's luminosity distance to values that are at odds with CMB and BAO. This clearly shows how the  $\Lambda$CDM model cannot possibly solve the $M$ crisis \cite{Camarena:2021jlr,Efstathiou:2021ocp}.
The more flexible $w$CDM, CPL and PEDE models fare much better but still much worse than the $LwMT$ and $LMT$ models which can fit all observables well.

All the models are forced to be consistent with the local Cepheid-calibrated value $M_c$ at the 1$\sigma$ level. As a result, in order to achieve consistency with $M_c$ the values of the other parameters differ significantly from the relevant \lcdm values. Also, the imposed significantly higher ($M>-19.28$) than the best fit inverse distance ladder value ($M=-19.4$) forces the MCMC process to restrict the rest of the parameters, thus explaining the extremely low uncertainties in order to achieve the best possible quality of fit to the data. We also stress that the peak structure of the CMB in the damping tail constrains very well the combination $\Omega_mh^2$. Therefore, once we force $M$-$H_0$ to be in agreement with SH0ES, we need a lower value of $\Omega_m$ different from \lcdm to compensate for the higher $H_0$ value and keep the peak structure unaltered. Once we relax the $M$ prior (Table \ref{tab:bfall_gaussian}) $\Omega_m$ can go back to the \lcdm value.

\subsection{Dark energy models comparison using the local Cepheid prior on \boldmath{$M$}} \label{secgau}

Here, we adopt the local  Gaussian prior of Eq.~\eqref{Mprior}. The constraints on cosmological parameters are given in Table \ref{tab:bfall_gaussian}, while the corresponding $68.3 \%$--$95.5 \%$ confidence contours are shown in Fig.~\ref{fig:contours_Mgaussian}. 
The transition $LwMT$/$LMT$ models fare significantly better than the other models, providing an absolute magnitude that is consistent with the Cepheid calibration of Eq.~\eqref{Mprior}.
The $w$CDM and CPL models achieve a slightly better fit to data as compared to $\Lambda$CDM, while the PEDE model has a significantly worse fit to data, in agreement with previous findings \cite{Pan:2019hac}. Note that constraints for the $LMT$ and $LwMT$ models have been not included in Fig. \ref{fig:contours_Mgaussian} for the sake of clarity. Constraints for these model are instead showed in Fig. \ref{fig:contours_Mgaussian_2} of the Appendix~\ref{appB}.

\subsection{Model selection}

To select the best model one cannot just look at the quality of fit but it is essential to include the information on the number of parameters and their priors. Here, we only consider the case of Section~\ref{secgau} as it uses the actual Cepheid prior of Eq.~\eqref{Mprior} from SH0ES.
We adopt two approaches. First, we consider the Akaike Information Criterion (AIC) \cite{AkaikeCrit,Nesseris:2012cq}, defined as
\be
AIC \equiv -2 \, \ln \mathcal{L}_\textrm{max}+2\, N_\textrm{tot}=\chi_\textrm{min}^2 +2\, N_\textrm{tot} \,, \label{eq:AIC}
\ee
where $N_\textrm{tot}$ corresponds to the total number of free parameters of the considered model and $\mathcal{L}_\textrm{max}$ corresponds to the maximum likelihood.
This criterion penalizes a model for any extra parameters. Using Eq.~\eqref{eq:AIC} 
we calculate the AIC values for all the models of Table \ref{tab:bfall_gaussian} and  construct the corresponding differences $\Delta AIC \equiv AIC_\textrm{model}-AIC_\textrm{$\Lambda$CDM}$, see Table  \ref{tab:statcritgau}.
If $|\Delta AIC| \leq 2$, then the compared models can be interpreted as consistent with each other, while if $|\Delta AIC| \geq 4$ it is an indication that the model with the larger AIC value is disfavored~\cite{Nesseris:2012cq}. We can see that the $LwMT$/$LMT$ models are strongly favored over \lcdmnospace, and that the PEDE model is strongly disfavored.

We also use the \href{https://github.com/yabebalFantaye/MCEvidence/}{MCEvidence} package \cite{Heavens:2017afc} in order to compute the Bayesian evidences (marginal likelihoods) of each model of Table \ref{tab:bfall_gaussian} using their respective MCMC chains. This algorithm obtains the posterior for the marginal likelihood, using the $k$-th nearest-neighbour Mahalanobis distances \cite{mahalanobis1936generalized} in the parameter space. In our analysis we have adopted the $k=1$ case to minimize the effects of the inaccuracies associated with larger dimensions of the parameter space and smaller sample sizes. The strength of the evidence presented in favor or against a model in a comparison, can be found using the revised Jeffreys' scale \cite{Trotta:2008qt}. Specifically, in a comparison between two models via the Bayes factor $B$ (ratio of evidences), if $|\ln B|  < 1$ the models are comparable with none of them being favored, for $1 < | \ln B|  < 2.5$ one model shows weak evidence in its favor, if $2.5 < | \ln B|  < 5$ the model in question has moderate evidence on its side, and lastly in the case of $| \ln B|> 5$ one model is strongly favored over the other.
From Table \ref{tab:statcritgau} one can see that PEDE is strongly disfavored, $w$CDM and CPL weakly favored and disfavored, respectively, and that the $LwMT$/$LMT$ models are strongly favored over \lcdmnospace.

\section{Discussion and conclusions}
\label{sec:concl}

We have investigated the quality of fit to cosmological data of five models that attempt to solve the $H_0$ crisis. Besides the standard \lcdm model, we considered three smooth $H(z)$ deformation models ($w$CDM, CPL and PEDE) and two models that allow for a sudden transition of the SnIa absolute magnitude $M$ at a recent cosmological redshift $z_t$.
We performed model selection via the Akaike Information Criterion and the Bayes factor. This is a more detailed and extended fit to the data that includes the full CMB angular power spectrum, instead of just the peak locations, discussed in the previous studies \cite{Alestas:2020zol, Marra:2021fvf} that introduced the ultra-late transition idea. We have also included additional cosmological models to compare the fit with the $M$ transition models and implemented different priors and model selection criteria.

We found that the transition models are strongly favored  with respect to the \lcdm model. We also found that PEDE is strongly disfavored and that $w$CDM and CPL are weakly favored and disfavored, respectively.
Specifically, only $M$-transition models are able to maintain consistency with the SnIa absolute magnitude $M_c$ measured by Cepheid calibrators while at the same time maintaining a quality of fit to the cosmological data at $z>0.01$ that is identical with that of $\Lambda$CDM. 

\begin{centering}
\begin{table*}[t]
\caption{Constraints at 68\% CL of the basic parameters for all the considered dark energy models, including the SH0ES measurement $H_0 = 73.2 \pm 1.3 \, \mathrm{km}\,\mathrm{s}^{-1}\,\mathrm{Mpc}^{-1}$ \cite{Riess:2020fzl}. Clearly all the considered dark energy models except the $LwMT$ model with $z_t \geq 0.01$ and $LMT$ with $z_t=0.01$ give a SnIa absolute magnitude $M$ that is inconsistent with the local calibration $M_c=-19.24 \pm 0.04 \, \mathrm{mag}$ of Eq.~\eqref{Mprior}. However, this statistical inconsistency is not included in the $\chi^2$ that is used to interpret the results, see discussion in \citep{Camarena:2021jlr}.}
\label{tab:bfallwRiess}
\resizebox{\textwidth}{!}{
\begin{tabular}{|c|c|c|c|c|c|c|}
 \hline 
 \rule{0pt}{3ex}  
  Parameters & \lcdm & $w$CDM & CPL & $LwMT$ & PEDE & $LMT$\\
  & & & & $(z_t \geq 0.01)$ & & $(z_t=0.01)$\\
    \hline
    \rule{0pt}{3ex}  
$\Omega_\mathrm{m,0}$  & $0.3022_{-0.0052}^{+0.0050}$ & $0.2967_{-0.0064}^{+0.0067}$ & $0.2951_{-0.0067}^{+0.0063}$ & $0.2989_{-0.0060}^{+0.0055}$ & $0.281\pm 0.005$ & $0.3021_{-0.0052}^{+0.0053}$\\ 
$n_s$  & $0.9705 \pm 0.0037$ & $0.9684\pm 0.004$ & $0.9668 \pm 0.0040$ & $0.9706\pm 0.0037$ & $0.9621_{-0.0034}^{+0.0036}$ & $0.9705\pm 0.0038$\\   
$H_0$ & $68.36 \pm 0.4$ & $69.17_{-0.76}^{+0.65}$ & $69.50 \pm 0.71$ & $68.71\pm 0.5$ & $71.69_{-0.46}^{+0.45}$ & $68.36_{-0.41}^{+0.40}$\\
$\sigma_\mathrm{8,0}$  & $0.8075_{-0.0064}^{+0.0058}$ & $0.8183_{-0.01}^{+0.0089}$ & $0.8258 \pm 0.0099$ & $0.8098\pm 0.0064$ & $0.8531_{-0.0058}^{0.0064}$ & $0.8086_{-0.0064}^{+0.0058}$\\
\rowcolor{LightGreen}
$M$ & $-19.40\pm 0.01$ & $-19.38 \pm 0.02$ & $-19.37_{-0.018}^{+0.017}$ & $-19.24$ & $-19.34 \pm 0.01$& $-19.24$\\
\hline
\cellcolor{LightGreen}$\Delta M$  & - & - & - & \cellcolor{LightGreen}$-0.1652\pm 0.011$ & -  & \cellcolor{LightGreen}$-0.159\pm 0.011$\\
\cellcolor{LightGreen}$M_> \equiv M_c+\Delta M$ & - & - & - & \cellcolor{LightGreen}$-19.405 \pm 0.011$ & - & \cellcolor{LightGreen}$-19.40 \pm 0.011$\\
$\Delta w$  & - & - & - & $>-0.7$ & - & -\\
$a_t$  & - & - & - & $>0.98$ & - & -\\
$w_0$  & - & $-1.038_{-0.018}^{+0.031}$ & $-0.9576_{-0.078}^{+0.075}$ & - & - & -\\ 
$w_a$  & - & - & $-0.38_{-0.27}^{+0.32}$ & - & - & -\\
\hline
 \rule{0pt}{3ex}  
$\chi^2$  & $3849$  & $3846$ & $3845$ & $3846$ & $3862$ & $3850$\\  
\rowcolor{LightCyan}
$\Delta \chi^2$ & - & $-3$ & $-4$ & $-3$ & $+13$ & $+1$\\
\hline
\end{tabular}}
\end{table*}
\end{centering}

The required transition with magnitude $\Delta M$ can be induced by a corresponding transition of the effective gravitational constant $G_{\rm eff}$ which determines the strength of the gravitational interactions \cite{Marra:2021fvf}.  The corresponding magnitude of the $G_{\rm eff}$ transition depends on the power value $b$ of the expression that connects the evolving Newton's constant $G_{\rm eff}$ with the  absolute luminosity $L$ of a SnIa:
\be
L \sim G_{\rm eff}^b \,. \label{eq:abslumGeff}
\ee
In the case of the $LwMT$ and $LMT$ transition models the transition in $M$ implies a transition in $\mu \equiv G_{\rm eff}/G_{\rm N}$. In particular, for $z>z_t$, it is:
\be 
\mu= 1+ \frac{\Delta G_{\rm eff}}{G_{\rm N}} \equiv 1+\Delta \mu, \label{eq:mudeltamu}
\ee
while for $z<z_t$ we have $\mu=1$.    Since, $\Delta \mu \ll 1$, we can assume without loss of generality that $\ln(1+\Delta \mu) \simeq \Delta \mu$, so it is straightforward to show that Eq.~\eqref{eq:abslumGeff} corresponds to
\be 
\Delta M=-\frac{5b}{2} \frac{\ln \mu}{\ln 10 }. \label{eq:dmeq1}
\ee
Using the definition \eqref{eq:mudeltamu}, for $z>z_t$, we have
\be 
\ln \mu \simeq \Delta M. \label{eq:dmeq2}
\ee
Therefore, substituting \eqref{eq:dmeq2} in \eqref{eq:dmeq1} and solving with respect to $b$, we derive
\be 
b=-\frac{2 \ln 10}{5} \frac{\Delta M}{\Delta \mu} \,. \label{eq:bform}
\ee
We can constrain $b$ based on Eq.~\eqref{eq:bform} and the fact that it obeys the general bounds $|b| \in [b_{\rm min},+\infty)$ (the $+\infty$ corresponds to the \lcdm/GR case where $\Delta \mu=0$). Taking the absolute value of Eq.~\eqref{eq:bform} and setting from Table \ref{tab:LMPTbf} the 2-$\sigma$ upper bound $|\Delta M|_{\rm min}=-0.172+2 \times 0.012=-0.148$ mag and $|\Delta \mu|_{\rm min}=0.05$ \cite{Alvey:2019ctk}, a measurement obtained using up to date primitive element abundances, cosmic microwave background as well as nuclear and weak reaction rates, $b_{\rm min}$ assumes the following $2\sigma$ range 
\be 
b_{\rm min,0.05}=(-\infty,-2.7] \cup [2.7,+\infty).
\ee
Similarly, if we consider the constraint from the Hubble diagram SnIa \cite{Gaztanaga:2008xz}, a measurement derived using luminous red galaxies, as well as from Paleontology \cite{Uzan:2002vq}, a measurement obtained using the age of bacteria and algae, that indicate $|\Delta \mu|_{\rm min}=0.1$, we derive
\be 
b_{\rm min,0.1}=(-\infty,-1.4] \cup [1.4,+\infty). 
\ee
This range includes the simple expectation that emerges if we assume that the SnIa absolute luminosity is proportional to the Chandrasekhar mass $L\sim M_\mathrm{Ch}\sim G_{\rm eff}^{-3/2}$ which leads to $b=-3/2$.
    
If the $M$ transition is due to a gravitational transition with a lower value of $G_{\rm eff}$ at $z>z_t$ then this class of models also has the potential to address the growth tension as discussed in previous studies \cite{Marra:2021fvf}. It should also be stressed that such a gravitational transition would be consistent with solar system tests of modified gravities without the need for screening since the value of $G_{\rm eff}$ is predicted to be constant at $z<z_t$ and therefore no modification of the planetary orbits is expected since the time these orbits have been monitored. However, at the time of the gravitational transition (about $100 \mathrm{Myrs}$ ago) a disruption of the planetary orbits and comets is expected \cite{Perivolaropoulos:2022vql}. Such a prediction may be consistent with the observational fact that the rate of comets that hit the Earth and the Moon has increased by a factor of 2-3 during the past $100 \mathrm{Myrs}$ \cite{1998JRASC..92..297S, 1995hdca.book.....G, https://doi.org/10.1029/97JE00114, https://doi.org/10.1029/1999JE001160, Ward2007TerrestrialCC}.

As discussed in, \textit{e.g.}, Refs. \cite{Livio:2018rue,Jha:2019svc,Flors:2019hdn}, SnIa progenitors are not necessarily Chandrasekhar-mass white dwarfs and a significant fraction can arise from sub-Chandrasekhar explosions. While this surely calls for a more detailed analysis of the dependence of the SnIa luminosity on $G_{\rm eff}$, one must note that the Chandrasekhar mass scale is a fundamental reference scale that plays an important role in all SnIa explosions. However, a nontrivial relation between progenitors and SnIa could again imply that the relation of Eq. \eqref{eq:abslumGeff} could feature a value of $b$ different from $-3/2$ corresponding to the simplest case where $L\sim M_{Chandra}$. 

\begin{figure*}[t!]
\centering
\includegraphics[width =  \textwidth]{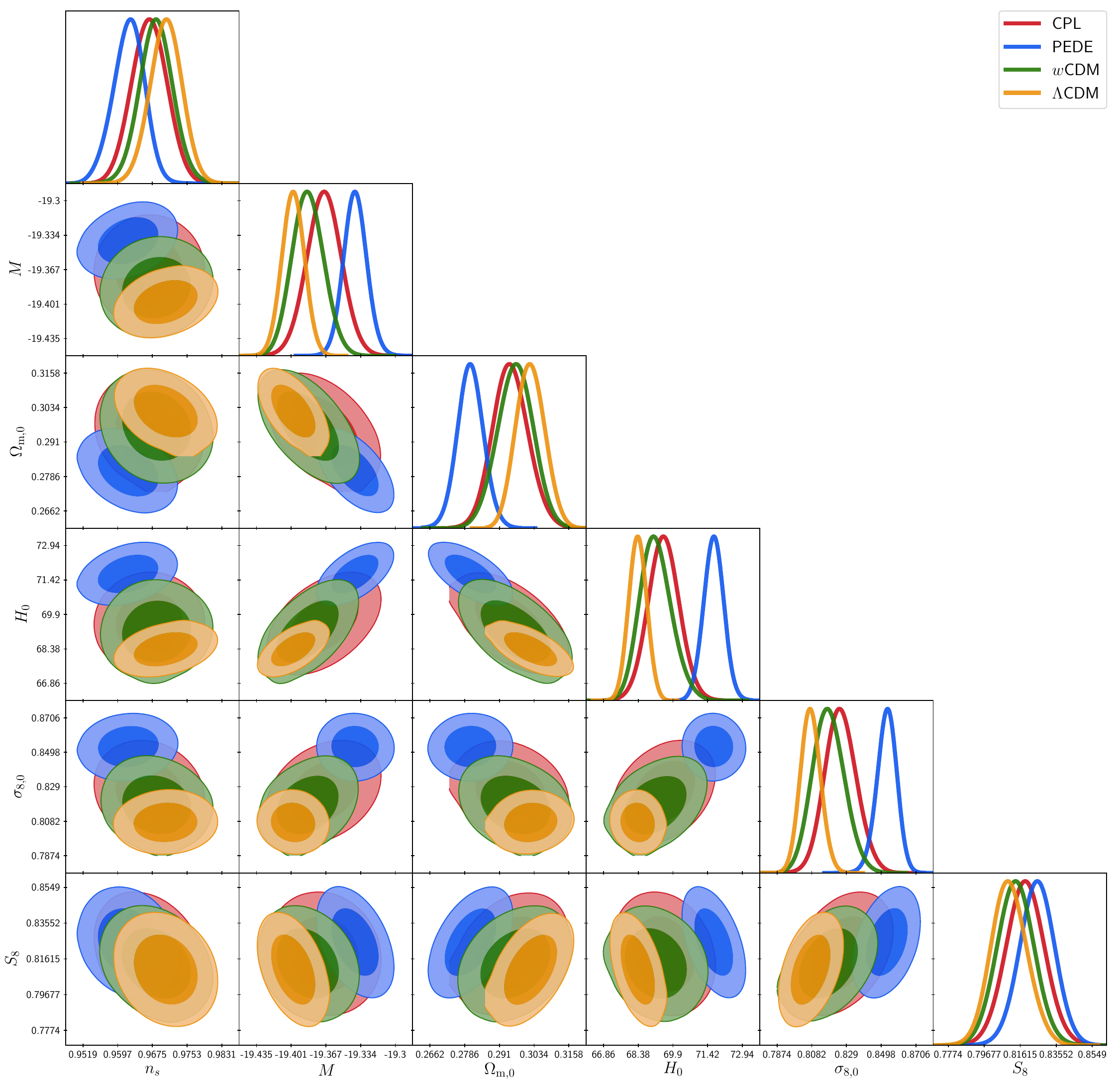}
\caption{The $68.3 \%$--$95.5 \%$ confidence contours for the common parameters of the $\Lambda$CDM, CPL, $w$CDM  and PEDE  dark energy models corresponding to the constraints given in Table \ref{tab:bfallwRiess}. We used the CMB+BAO+Pantheon+RSD likelihoods, including the SH0ES measurement $H_0 = 73.2 \pm 1.3 \, \mathrm{km}\,\mathrm{s}^{-1}\,\mathrm{Mpc}^{-1}$ \cite{Riess:2020fzl}. The $M$ value of the models is inconsistent with the local calibration of Eq.~\eqref{Mprior} $(M_c=-19.24 \pm 0.04 \, \mathrm{mag})$.} 
\label{fig:contours_Mpr2}
\end{figure*} 

\begin{figure*}[ht!]
\centering
\includegraphics[width = \textwidth]{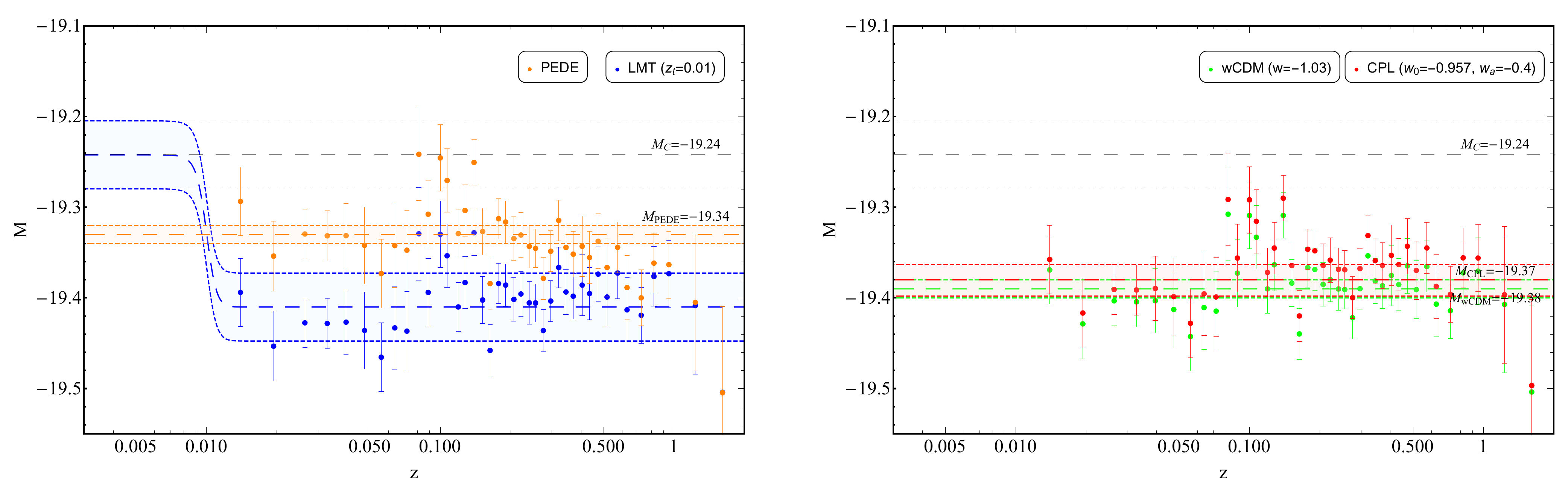}
\caption{The best fit absolute magnitude $M$ of the binned Pantheon data as a function of the redshift $z$. In the left panel we show the corresponding best fit data for the $LMT$ model with $z_t=0.01$ (blue points) and PEDE models (orange points). In the right panel we show the corresponding best fit data for the $wCDM$ model with $w=-1.03$ (green points) and CPL models (red points). Clearly, all the models provide a value that is inconsistent with the measured  Cepheid absolute magnitude $M_c$ (straight dashed lines), unless a model with a transition on $M$ (such as $LMT$) is considered.} 
\label{fig:mzplt}
\end{figure*}

Therefore, interesting extensions of our analysis include the following:
\begin{itemize}
    \item 
    The search for traces or constraints of a gravitational transition in geological, solar system and astrophysical data.
    \item
    The construction of simple theoretical modified gravity models that can naturally induce the required transition of the effective Newton's $G_{\rm eff}$ at low redshifts ($z_t<0.01$) perhaps avoiding fine tuning issues.
    \item
    The possible identification of alternative non-gravitational physical mechanisms that could induce the transition of SnIa at low redshifts.
    \item
    The search for systematic effects in the Cepheid data and parameters that could mimic such a transition and/or induce a higher value of $M$ for SnIa than the one currently accepted.
\end{itemize}

In conclusion the $M$-transition class of models is an interesting new approach to the Hubble and possibly to the growth tension that deserves further investigation.

\ 

\textbf{Numerical Analysis Files}: The numerical files for the reproduction of the figures can be found in the GitHub repository \href{https://github.com/lkazantzi/H0_Model_Comparison}{H0\_Model\_Comparison}  under the MIT license.

\section*{Acknowledgements}

The MCMC chains were produced in the Hydra cluster at the Instituto de F\'isica Te\'orica (IFT) in Madrid and in the CHE cluster, managed and funded by COSMO/CBPF/MCTI, with financial   support   from   FINEP   and   FAPERJ, and  operating  at  the  Javier  Magnin  Computing  Center/CBPF, using \href{https://github.com/brinckmann/montepython_public}{MontePython}/\href{https://github.com/lesgourg/class_public}{CLASS} \cite{Brinckmann:2018cvx, Audren:2012wb, Blas_2011}.
GA's research was supported by the project “Dioni: Computing Infrastructure for Big-Data Processing and Analysis” (MIS No. 5047222) co-funded by European Union (ERDF) and Greece through Operational Program “Competitiveness, Entrepreneurship and Innovation”, NSRF 2014-2020.
DC thanks CAPES for financial support.
EDV is supported by a Royal Society Dorothy Hodgkin Research Fellowship.
LK is co-financed by Greece and the European Union (European Social Fund- ESF) through the Operational Programme ``Human Resources Development, Education and Lifelong Learning" in the context of the project ``Strengthening Human Resources Research Potential via Doctorate Research – 2nd Cycle" (MIS-5000432), implemented by the State Scholarships Foundation (IKY). 
VM thanks CNPq (Brazil) and FAPES (Brazil) for partial financial support. This project has received funding from the European Union’s Horizon 2020 research and innovation programme under the Marie Skłodowska-Curie grant agreement No 888258.
SN acknowledges support from the Research Project PGC2018-094773-B-C32, the Centro de Excelencia Severo Ochoa Program SEV-2016-0597 and the Ram\'{o}n y Cajal program through Grant No. RYC-2014-15843.

\appendix

\begin{figure*}[t!]
\centering
\includegraphics[width = \textwidth]{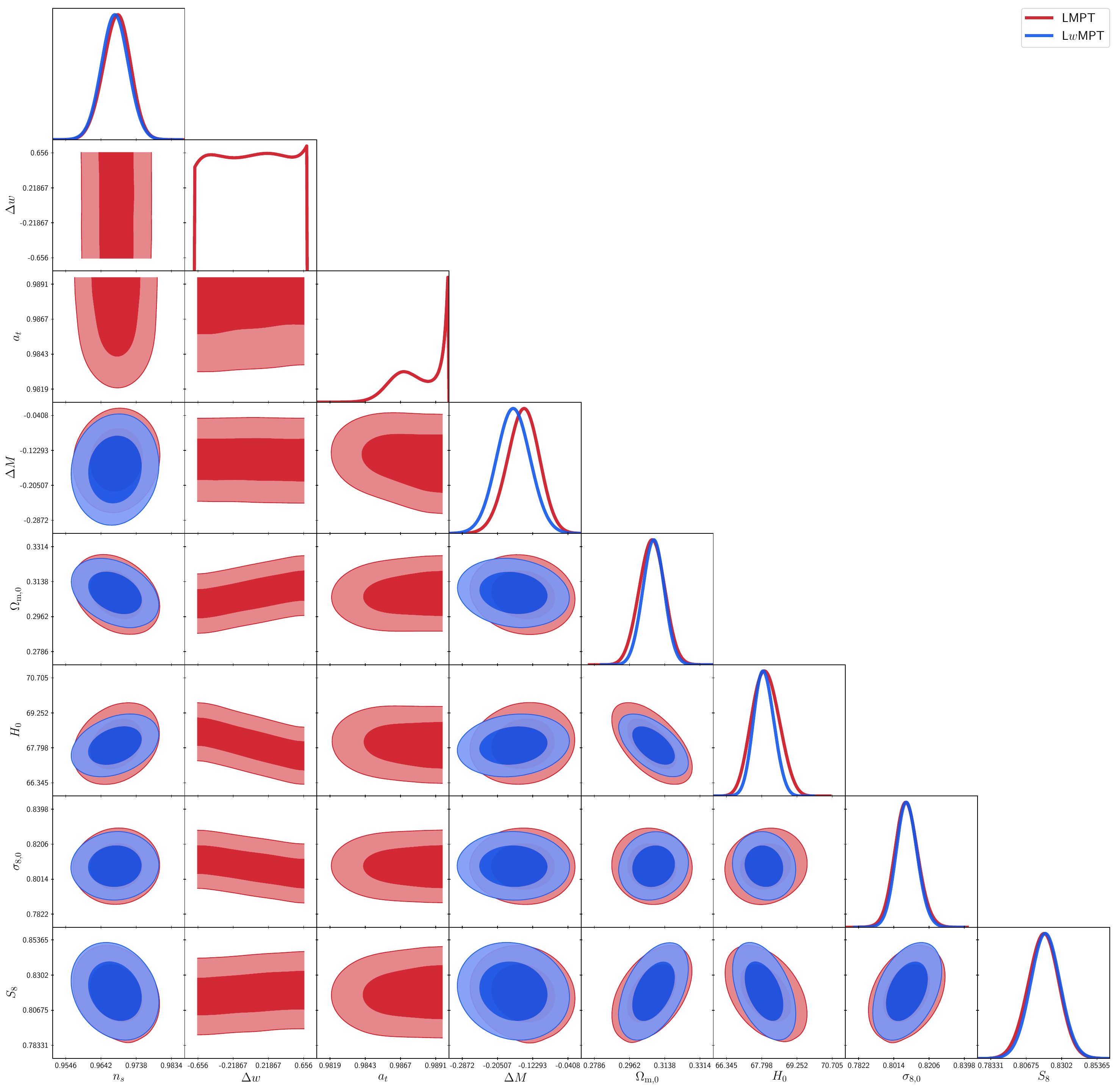}
\caption{The $68.3 \%$--$95.5 \%$ confidence contours of the $LMT$, and $LwMT$ dark energy models with the  prior $ M = -19.24 \pm 0.04$ mag of Eq.~\eqref{Mprior} from  SH0ES (corresponding bounds in Table \ref{tab:bfall_gaussian}).} 
\label{fig:contours_Mgaussian_2}
\end{figure*} 

\begin{figure*}[t!]
\centering
\includegraphics[width = \textwidth]{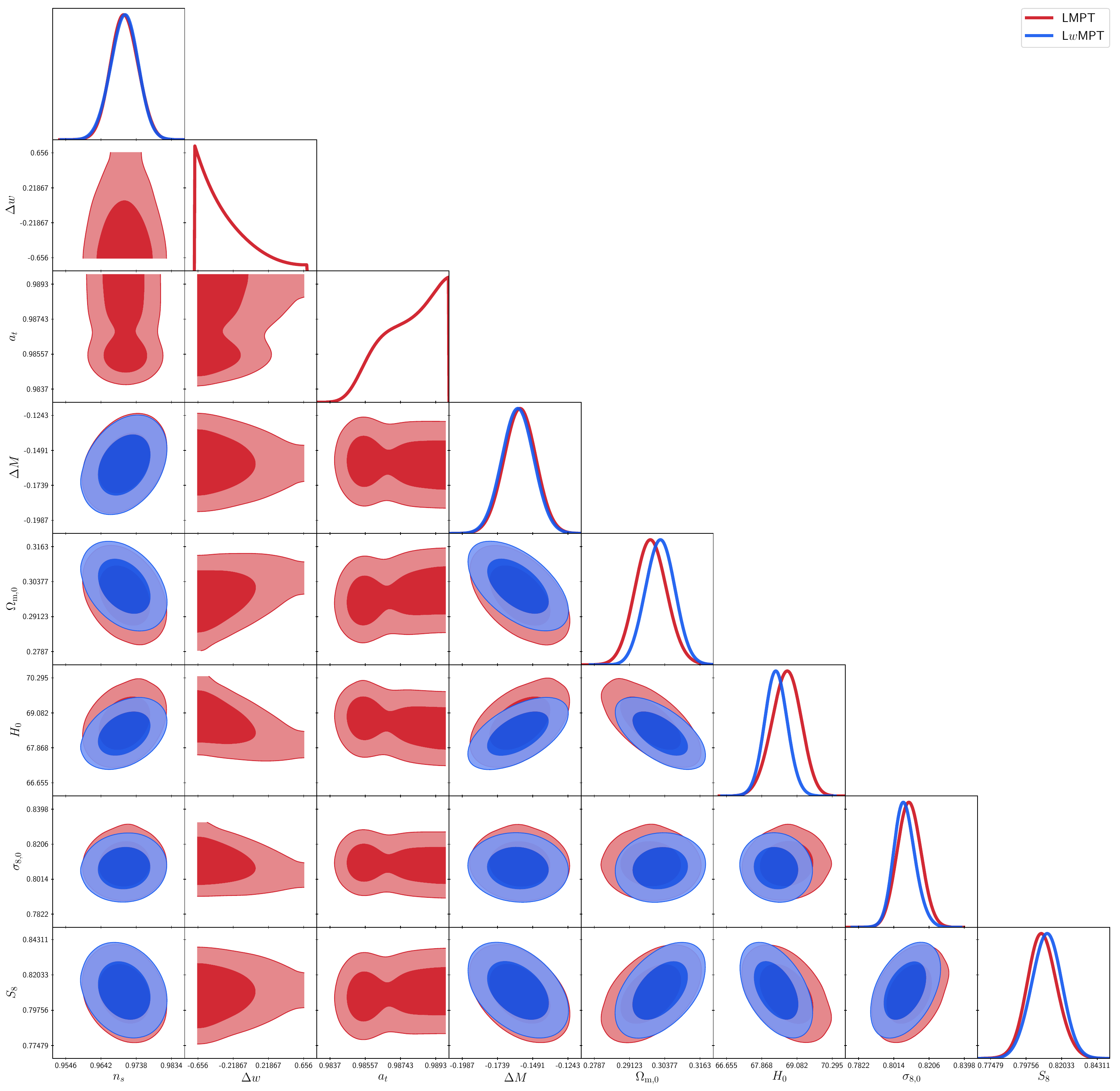}
\caption{The $68.3 \%$--$95.5 \%$ confidence contours for the common parameters of the $LMT$, and $LwMT$ dark energy models corresponding to the constraints given in Table \ref{tab:bfallwRiess}. We used the CMB+BAO+Pantheon+RSD likelihoods, including the SH0ES measurement $H_0 = 73.2 \pm 1.3 \, \mathrm{km}\,\mathrm{s}^{-1}\,\mathrm{Mpc}^{-1}$ \cite{Riess:2020fzl}.} 
\label{fig:contours_H0_2}
\end{figure*} 

\section{Analysis of the Dark Energy Models including the Local $H_0$ Measurement}
\label{appA}

We  repeat the analysis for the models in question including the latest SH0ES measurement, $H_0 = 73.2 \pm 1.3 \, \mathrm{km}\,\mathrm{s}^{-1}\,\mathrm{Mpc}^{-1}$ \cite{Riess:2020fzl}, instead of the local prior on $M$ of Eq.~\eqref{Mprior} that was adopted in Section~\ref{secgau}. This is done in order to show that, despite the strong constraining nature of the SH0ES measurement, the obtained absolute magnitude $M$ for smooth $H(z)$ deformation models are inconsistent with the measured Cepheid absolute magnitude $M_c=-19.24 \pm 0.04 \, \mathrm{mag}$ of Eq.~\eqref{Mprior}.
It is worth stressing that it is preferable to adopt the local prior on $M$ for the following reasons \citep{Camarena:2021jlr}: i) one avoids double counting low-$z$ supernova, ii) the statistical information on $M$ is included in the analysis, iii) one avoids adopting a low-$z$ cosmography, with possibly wrong parameters, in the analysis. 

Repeating the MCMC analysis and using the same likelihoods described in Section \ref{sec:lwmptdat}, we obtain the constraints on cosmological parameters for all the models as shown in Table \ref{tab:bfallwRiess}. The corresponding $68.3 \%$--$95.5 \%$ confidence contours of the common parameters of the models are illustrated in Fig.~\ref{fig:contours_Mpr2}. Constraints for the $LMT$ and $LwMT$ models have been not included in Fig. \ref{fig:contours_Mpr2}, instead we  show those constraints in Fig. \ref{fig:contours_H0_2}.

Clearly, all of the considered models (except the models with transitions) tend to prefer a significantly lower value for $M$ (which is considered to be constant) compared to $M_c$. This is also evident in Fig.~\ref{fig:mzplt}, where the best fit absolute magnitude $M$ of the binned Pantheon data is shown. In particular, in the case where no prior on $M$ is imposed, two of the considered models, \ie $w$CDM and CPL, produce a $H_0$ best fit value that is inconsistent with the SH0ES measurement \cite{Riess:2020fzl} at more than $2.4\sigma$. Regarding the $LwMT$ with $(z_t \geq 0.01)$, even with the SH0ES measurement, the best fit value of $a_t$ parameter remains unaffected, continuing to favor a transition at very low redshifts. Conclusively, even though the majority of dark energy models discussed in this work (except PEDE and $LMT$) display a better quality of fit to the data than that of \lcdm (cyan row of Table \ref{tab:bfallwRiess}), they fail to give an $M$ value consistent with the $M_c$ measurement (except the $LwMT$ and $LMT$ models) despite the fact that some of them (such as PEDE) provide a $H_0$ measurement, that is consistent with the SH0ES measurement at the $1\sigma$ level.

\section{Contours plots for the $LMT$ and $LwMT$ models}
\label{appB}

Here, we show contours plots relative to analyses with the models $LMT$ and $LwMT$. Fig. \ref{fig:contours_Mgaussian_2} shows the $68.3 \%$--$95.5 \%$ confidence contours of cosmological parameters of models $LMT$ and $LwMT$ for the analysis with the Gaussian prior on $M$. On the other hand, Fig. \ref{fig:contours_H0_2} shows the $68.3 \%$--$95.5 \%$ confidence contours of cosmological parameters of models $LMT$ and $LwMT$ for the analysis that includes a Gaussian prior on $H_0$.

\raggedleft
\bibliography{Bibliography}

\end{document}